\documentclass[aps,physrev,reprint,superscriptaddress]{revtex4-2}

\usepackage{amsmath} 
\usepackage{graphicx}
\usepackage{xcolor}
\usepackage{dcolumn}
\usepackage{bm}
\usepackage{placeins}
\usepackage{array}
\usepackage{siunitx}
\usepackage{orcidlink}

\begin{document}


\title{Mechanical acclimation of bacterial growth and division across generations} 

\author{Chenyu Jin \,\orcidlink{0000-0002-5552-0340}}
 \email{Contact author: chenyujin@fudan.edu.cn}
 \affiliation{
 Physics of Living Matter, Department of Physics and Materials Science, University of Luxembourg, \\
 162 A, Avenue de la Faïencerie, L-1511 Luxembourg City, Luxembourg.
}%
\affiliation{ 
  Yiwu Research Institute of Fudan University, Zhejiang 322000, People's Republic of China
}%
\affiliation{ 
  International Institute for Intelligent Nanorobots and Nanosystems, Fudan University, Shanghai 200438, People's Republic of China
}

\author{G. H. Philipp Nguyen \,\orcidlink{0000-0003-0304-1441}}%
\affiliation{%
 Institut f\"{u}r Theoretische Physik II: Weiche Materie,
    Heinrich-Heine-Universit\"{a}t D\"{u}sseldorf, \\
    Universit\" {a}tsstraße 1, 40225, D\"{u}sseldorf, Germany.
}%

\author{Ren\'{e} Wittmann \,\orcidlink{0000-0002-8787-1823}}%
\affiliation{Institut f\"ur Tierische Lebensmittel, Max Rubner-Institut, 95326 Kulmbach, Germany}
\affiliation{%
 Institut f\"{u}r Theoretische Physik II: Weiche Materie,
    Heinrich-Heine-Universit\"{a}t D\"{u}sseldorf, \\
    Universit\" {a}tsstraße 1, 40225, D\"{u}sseldorf, Germany.
}%

\author{Hartmut L\"{o}wen \,\orcidlink{0000-0001-5376-8062}}%
\affiliation{%
 Institut f\"{u}r Theoretische Physik II: Weiche Materie,
    Heinrich-Heine-Universit\"{a}t D\"{u}sseldorf, \\
    Universit\" {a}tsstraße 1, 40225, D\"{u}sseldorf, Germany.
}%

\author{Anupam Sengupta \,\orcidlink{0000-0001-5592-7864}}%
 \email{Contact author: anupam.sengupta@uni.lu}
\affiliation{
 Physics of Living Matter, Department of Physics and Materials Science, University of Luxembourg, \\
 162 A, Avenue de la Faïencerie, L-1511 Luxembourg City, Luxembourg.
}%
\affiliation{
 Institute for Advanced Studies, University of Luxembourg, \\
 2 Avenue de l’Université, L-4365, Esch-sur-Alzette, Luxembourg.
}%
\affiliation{
 Institute of Microbiology, Department for Environment, Constructions and Design, University of Applied Sciences and Arts of Southern Switzerland (SUPSI), \\
 Via Flora Ruchat-Roncati 15, 6850, Mendrisio, Switzerland.
}%

\date{\today}

\begin{abstract}

Bacteria frequently encounter mechanical confinement in natural and engineered environments, yet how an abrupt change in mechanical conditions reorganizes growth and division across generations remains poorly understood. Using single-cell time-lapse microscopy, we tracked \textit{Escherichia coli} across generations following transfer from liquid culture to glass-attached or glass–gel-confined conditions. This mechanical confinement causes an immediate reduction in elongation rate, followed by a multi-generational mechanical acclimation process whose dynamics depend systematically on confinement strength. During this relaxation, division behavior becomes more heterogeneous and cells temporarily depart from canonical \textit{adder}-like size regulation, exhibiting \textit{timer}-like correlation signatures. Using a stochastic cell-cycle model that resolves the generation-dependent correlation signatures of different size-control strategies, we show that changes in elongation rate alone cannot account for the observed transition in growth--division coordination. Despite this transient reorganization, the birth-size distribution remains bounded and largely preserves its shape throughout acclimation. Notably, \textit{adder}-like regulation is restored before the elongation rate fully recovers, revealing distinct relaxation timescales for growth and size-control dynamics. These results establish bacterial size control under mechanical confinement as a dynamical nonequilibrium process in which growth and division can transiently reorganize while global size homeostasis is maintained.

\end{abstract}


\maketitle

\section{Introduction}
\noindent
Microbes thrive throughout the biosphere, from the atmosphere to deep oceans and subsurface rocks \cite{flemming2019bacteria, bar2018biomass, jin2024microbes}. Many of their habitats are spatially heterogeneous and mechanically confined, including soils \cite{ebrahimi2014, kondakova2013study, philippot2023}, depleted gas reservoirs \cite{heinemann2021, muhammed2022}, host tissues and mucus-rich environments \cite{kirch2012, wheeler2019, bansil2018, wu2023mucin}, and marine snow and biologically-active sediments \cite{dang2016, alcolombri2021sinking, borer2023, alcolombri2025biogel}. In such environments, bacterial cells experience combinations of compression, adhesion, steric restriction, and crowding, all of which can influence microbial growth and survival \cite{persat2015mechanical, dufrene2020mechanomicrobiology}. Understanding how single cells respond to these mechanical constraints is therefore essential for connecting microbial ecology with quantitative biophysics.

Bacteria can sense and respond to mechanical cues across scales, from changes in single-cell physiology to the collective organization of growing populations. At the single-cell level, surface attachment and external mechanical stress—such as hydrostatic pressure and soft-matter confinement—can alter cell metabolism \cite{kimkes2020bacteria,siryaporn2014surface, zhou2025confined}, change envelope mechanics \cite{pelech2025stress}, suppress cell elongation \cite{tuson2012measuring, wittmann2023collective}, and initiate biofilm formation \cite{chu2018self}. Intense mechanical constraints can further disrupt cell morphology and division \cite{kumar2013pressure, nepal2018dynamics, leblanc2024bacterial, luviano2026conditional, cayron2023bacterial}, leading to irregular cell shapes \cite{mannik2009bacterial, si2015bacterial}, delayed division, and in extreme cases, cell death \cite{manas2004morphological, kumar2013pressure}.
Beyond these single-cell responses, mechanical forces, confinement, and crowding can shape colony organization, spatial structure, topology, and collective behavior in growing microbial populations. \cite{allen2019bacterial, araujo2023steering, meacock2021bacteria, sengupta2020microbial, dhar2022self, wittmann2023collective, rani2025nascent, sreepadmanabh2024cell, hancock2024interplay, zimmer2026dimensionality}. These studies establish the mechanical environment as an active regulator of bacterial growth, division, morphology, and collective organization. However, majority of previous studies have characterized cells after they have already adapted to a sustained stressed or confined state, leaving the dynamics of the initial acclimation process largely unresolved.

Under steady growth conditions, \textit{E. coli} and \textit{Bacillus subtilis} exhibit robust size control in both microfluidic and agarose-pad assays \cite{jun2018fundamental, wang2010robust, campos2014constant, taheri2015cell, kennard2016individuality}. Bacterial size homeostasis relies on tight coupling between growth and division across generations. Single-cell measurements have shown that this coordination is often well described by the \textit{adder} principle, in which cells add an approximately constant size between birth and division, largely independent of their birth size \cite{campos2014constant, taheri2015cell}. Recent mechanistic studies have further suggested that division control can emerge from threshold accumulation of key division proteins, such as FtsZ, through balanced biosynthesis \cite{si2019mechanistic, miguel2025modulation}. These studies have focused primarily on cell-size control under steady-state growth. How this coordination is reshaped across generations following an abrupt switch in the mechanical environment remains poorly understood.

Here, we use single-cell time-lapse microscopy to track \textit{E. coli} across successive generations after the onset of mechanical confinement. We quantify elongation rates, division timing, and birth-size distributions and use generation-resolved correlations to characterize size control. 
We further develop a minimal stochastic model to compare the observed dynamics with idealized \textit{adder}-, \textit{sizer}-, and \textit{timer}-like reference behaviors. Our results show that confinement induces not only an immediate slowdown in growth, but also a transient transformation of canonical \textit{adder}-like growth--division coupling.

\section{Switch in mechanical environments initiates multi-generational growth acclimation}

\begin{figure}
	\centering
	\includegraphics[width=\columnwidth]{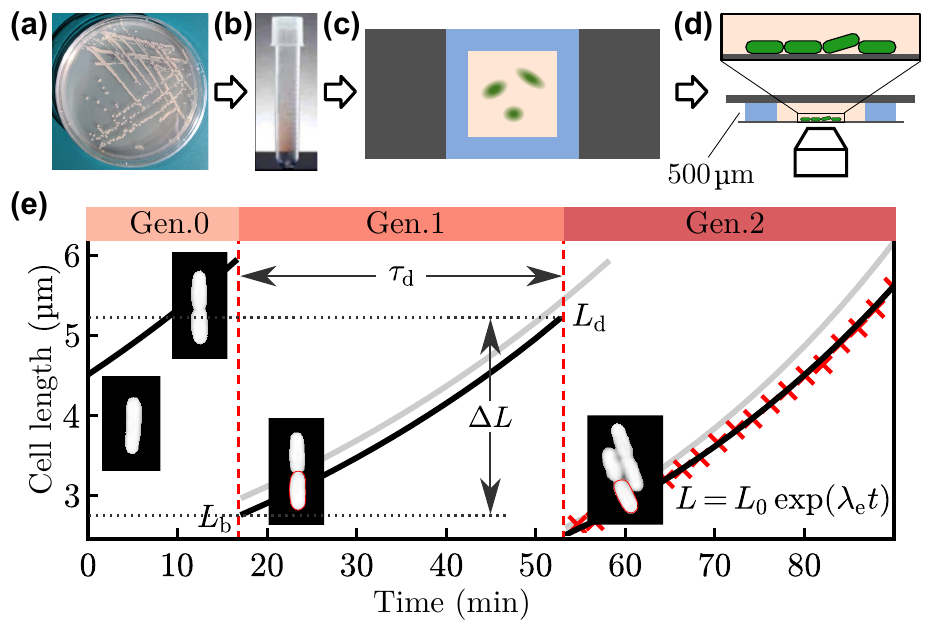} 
	\caption{\textbf{Growth-cycle metrics under single-cell confinement.}
\textbf{(a)--(d)} Experimental workflow for single-cell confinement assays. Exponentially growing \textit{E. coli} cells were transferred from liquid culture into confinement chambers for phase-contrast time-lapse imaging.
\textbf{(e)} Definitions of growth-cycle metrics. Cell-length trajectories of representative cells during the first three generations following confinement. Black curves show representative cell-length records during the first three generations following confinement.
Birth length, division length, added length, division time, division rate, and elongation rate are denoted by $L_\mathrm{b}$, $L_\mathrm{d}$, $\Delta L=L_\mathrm{d}-L_\mathrm{b}$, $\tau_\mathrm{d}$, $\lambda_\mathrm{d}=\ln2/\tau_\mathrm{d}$, and $\lambda_\mathrm{e}$, respectively. 
Elongation rates were obtained by fitting the measured cell lengths within each cell cycle to an exponential growth model; representative measurements used for fitting are indicated by red crosses.}
	\label{fig:setup} 
\end{figure}

To investigate how mechanical confinement influences bacterial growth and size regulation, exponentially growing \textit{E.~coli} cells were transferred from lysogeny broth (LB) liquid culture into three confinement conditions of increasing mechanical strength (Fig.~\ref{fig:setup}(a--d)): a glass-attached state in liquid LB medium (glass–liquid), confinement between glass and a $1.5\%$ agarose pad (glass–soft gel), and confinement between glass and a $6\%$ agarose pad (glass–hard gel). In all cases, colonies grew from isolated founder cells and expanded within chambers approximately \qty{500}{\micro\meter} thick. All confinement chambers were pre-equilibrated at $30^\circ\mathrm{C}$ to minimize temperature perturbation during transfer.

Single-cell growth and division were monitored by phase-contrast time-lapse microscopy at $2\,\mathrm{min}$ intervals for $3$--$7\,\mathrm{h}$ at $30^\circ\mathrm{C}$. Cell contours, lineages, and growth dynamics were extracted using automated segmentation and tracking pipelines, with manual corrections if necessary (see Supplementary Fig.~S1). Elongation rates were obtained by fitting the measured cell lengths within each cell cycle to an exponential growth model. Additional experimental and analysis details are provided in the Supplementary Material.

Following transfer into the confinement chamber, approximately $10$--$15\,\mathrm{min}$ was required to identify isolated founder cells and initiate time-lapse acquisition, which is shorter than a typical cell cycle ($\sim 36\,\mathrm{min}$ at $30^{\circ}$C). Cells present in the first recorded frame are therefore defined as generation $\mathrm{Gen.\,0}$, and their first division marks entry into $\mathrm{Gen.\,1}$.

For each cell cycle, the cell length immediately after division is defined as the birth length $L_\mathrm{b}$, and the length immediately before the subsequent division as the division length $L_\mathrm{d}$. The added length is $\Delta L = L_\mathrm{d} - L_\mathrm{b}$, and the division time is denoted by $\tau_\mathrm{d}$. The corresponding division rate is $\lambda_\mathrm{d}=\ln2/\tau_\mathrm{d}$. Single-cell elongation rates, $\lambda_\mathrm{e}$, were obtained by fitting cell-length trajectories to an exponential growth model (Fig.~\ref{fig:setup}(e)). Under ideal exponential growth with constant $\lambda_\mathrm{e}$ and $L_\mathrm{d} = 2L_\mathrm{b}$, the division and elongation rates satisfy $\lambda_\mathrm{d} = \lambda_\mathrm{e}$.
\bigskip

\begin{figure}
	\centering
	\includegraphics[width=\columnwidth]{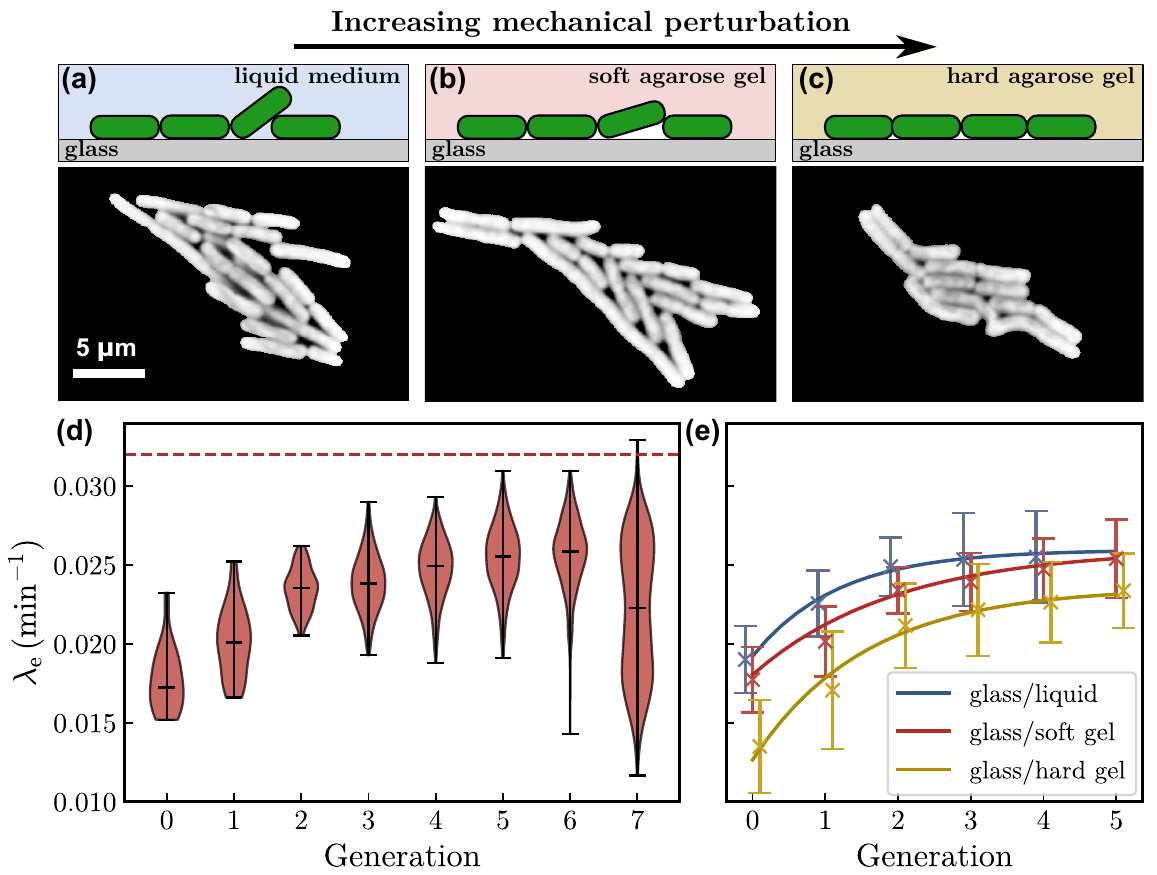} 
	\caption{\textbf{Growth acclimation after
    switch in mechanical environment.}
    \textbf{(a)--(c)} Experimental conditions of increasing confinement strength: \textbf{(a)} glass--liquid medium (glass-attached state in liquid LB medium), \textbf{(b)} glass--soft gel ($1.5\%$ agarose), and \textbf{(c)} glass--hard gel ($6\%$ agarose). The representative colony morphologies at 150 min, around $\mathrm{Gen.\,4}$.
    \textbf{(d)} Violin plots of single-cell elongation rate $\lambda_\mathrm{e}$ across generations following transfer into glass--soft gel confinement. Cells with outlier values beyond three standard deviations ($z$-score $>3$) were excluded. Black crosses mark the median, while vertical end markers denote the minimum and maximum values. Cells exhibit an initial reduction in elongation rate followed by gradual recovery over approximately $3$--$4$ generations. The mean elongation rate of dispersed cells in liquid culture is indicated by the red dashed line.
    \textbf{(e)} Recovery of elongation dynamics under different confinement strengths. All the elongation rates measured from $\mathrm{Gen.\,0}$ to $\mathrm{Gen.\,5}$ were fitted with the exponential recovery function $\lambda_\mathrm{e}(t)=\lambda_\infty-\Delta\lambda \,e^{-t/\tau}$ for cells grown under glass--liquid medium, glass--soft gel and glass--hard gel confinements. Mean $\lambda_\mathrm{e} \pm$ s.d. is shown for each generation, with data points from different confinement conditions slightly shifted along the $x$ axis slightly shifted to improve visual clarity.
 }
	\label{fig:ke-shock} 
\end{figure}

We observed a pronounced reduction in the single-cell elongation rate $\lambda_\mathrm{e}$ when exponentially growing \textit{E. coli} cells were transferred from liquid culture into three different confinement conditions (Fig.~\ref{fig:ke-shock}, Supplementary Fig.~S2). Dispersed cells in liquid culture elongated at an average rate of $\langle \lambda_\mathrm{e}^0 \rangle \approx 0.032\,\mathrm{min^{-1}}$, whereas cells in the initial confined generation ($\mathrm{Gen.\,0}$) elongated at reduced rates of approximately $0.019\,\mathrm{min^{-1}}$ in cells attached to glass in liquid LB medium (glass--liquid medium), $0.018\,\mathrm{min^{-1}}$ under glass--soft gel confinement and $0.013\,\mathrm{min^{-1}}$ under glass--hard gel confinement (Fig.~\ref{fig:ke-shock}(d,e)). 
Notably, cells continued to elongate exponentially after transfer, indicating that the confined environment remained permissive for growth. However, the substantial reduction in $\lambda_\mathrm{e}$ demonstrates that the transition into confinement, often regarded as a passive imaging condition in bacterial physiology experiments, represents an abrupt change in the mechanical environment that induces a pronounced physiological response, even in the absence of gel compression.

Across confinement conditions, elongation rates recovered approximately exponentially over multiple generations toward a plateau, consistent with a single-timescale acclimation process, $\lambda_\mathrm{e}(t)=\lambda_\infty-\Delta\lambda \,e^{-t/\tau}$. Increasing confinement strength generally reduced the recovered elongation rate $\lambda_\infty$ and increased the magnitude of the initial growth suppression $\Delta\lambda$. In contrast, the characteristic recovery timescale $\tau$ remained on the order of a few generations and did not show a clear monotonic dependence on confinement strength (see Supplementary Fig.~S3). The recovery was most pronounced during the first $3$--$4$ generations following confinement.

At later generations, such as $\mathrm{Gen.\,6}$ and $\mathrm{Gen.\,7}$, $\lambda_\mathrm{e}$ decreases again in a subset of cells, likely reflecting the collective mechano-response associated with increasing colony density \cite{wittmann2023collective}. 
Furthermore, stress buildup in the colony center can rotate cells and extrude them out of the monolayer \cite{allen2019bacterial, dhar2022self, you2018geometry, you2019}.
This mechanically driven monolayer-to-multilayer transition marks an early step toward biofilm-like colony architecture and can compromise reliable single-cell tracking. We therefore focus our analysis on the early acclimation phase preceding substantial multilayer growth.

A second hallmark of the confinement-induced stress response is filamentation \cite{kumar2013pressure, nepal2018dynamics, luviano2026conditional, cayron2023bacterial}, characterized by continued cell elongation despite delayed or impaired division. Beginning at $\mathrm{Gen.\,2}$, a subpopulation of cells ($\sim 7\%$ in area fraction, see Supplementary Fig.~S9) fails to divide within the expected cell cycle and continues to elongate while neighboring cells undergo multiple generations of division, before eventually exhibiting irregular division events. Representative examples are shown in Supplementary Fig.~S8(a--e). These observations indicate that cell elongation can continue despite substantial disruption of division under confinement.
 Similar filamentous phenotypes have previously been observed under diverse stress conditions and are generally associated with temporary suppression of division while growth continues. Additional analyses of filamentation, division-length distributions, and added-length statistics are provided in the Supplementary Figs.~S5--S9.

The emergence of filamentation demonstrates that mechanical confinement can transiently disrupt the coordination between growth and division during the acclimation phase.
This observation raises the question in how far also the growth--division coordination of nonfilamentous cells might be affected and whether size homeostasis remains preserved under mechanical confinement.
To this end, our further quantitative analysis was restricted to cells completing identifiable division cycles, while highly filamentous cells with unreliable contour extraction were excluded.

\FloatBarrier
\section{Birth-size statistics remain robust under mechanical confinements}

\begin{figure}
	\centering
	\includegraphics[width=\columnwidth]{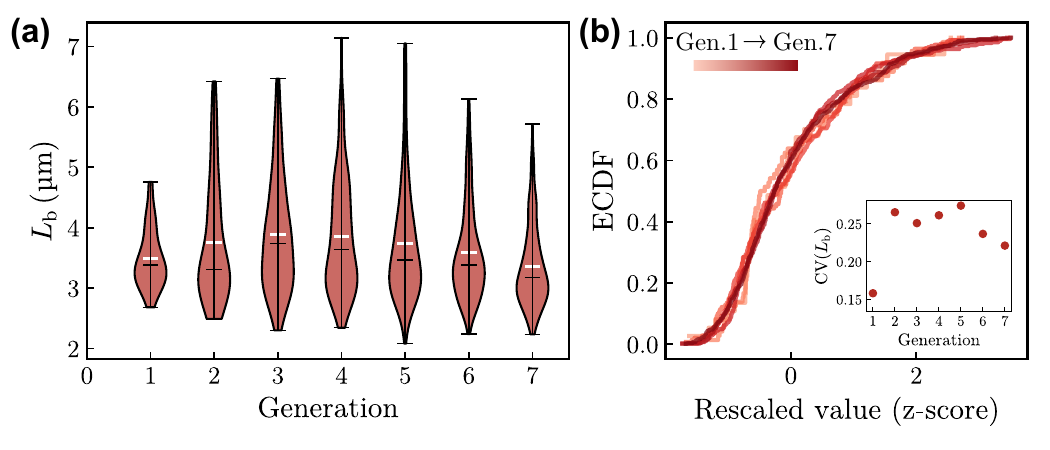} 
	\caption{\textbf{Birth-size statistics remain robust under mechanical confinement.}
    \textbf{(a)} Violin plots of birth length $L_b$ across generations following transfer into glass--soft gel confinement, showing that the characteristic birth size remains bounded despite the transient reduction in elongation rate. White bars denote the mean, black crosses the median, and vertical whiskers the distribution extrema. Analysis was restricted to cells completing identifiable division cycles; highly filamentous cells with unreliable contour extraction were excluded.
    \textbf{(b)} Empirical cumulative distribution functions (ECDFs) of $L_b$ after $z$-score normalization. The approximate collapse of the normalized ECDFs indicates that the standardized birth-length distributions retain a similar shape across generations. The normalization is defined as $z=(L_b-\mu)/\sigma$, where $\mu$ and $\sigma$ are the mean and standard deviation of $L_b$ for each generation, respectively.
    \textbf{Inset}: Coefficient of variation (CV) of $L_b$ across generations. Variability increases transiently during the acclimation phase before partially recovering at later generations.
    }
	\label{fig:homo} 
\end{figure}
 
To examine how bacterial cell size at birth responds to sudden mechanical confinement, we examined the distribution of birth lengths $L_\mathrm{b}$ across generations. Figure~\ref{fig:homo}(a) shows violin plots of $L_\mathrm{b}$ from $\mathrm{Gen.\,1}$ to $\mathrm{Gen.\,7}$ (sample sizes ranging from $37$ to $543$ cells per generation). 
Despite the pronounced reduction in elongation rate following confinement, the mean birth length remains approximately constant throughout the mechanical acclimation process.
The birth lengths generally follow a lognormal distribution (see Supplementary Fig.~S5). This is consistent with stochastic multiplicative fluctuations during exponential growth, moderated by different size-control mechanisms \cite{hosoda2011}.

Mechanical confinement nevertheless transiently increases the cell-to-cell variability. As shown in the inset of Fig.~\ref{fig:homo}(b), the coefficient of variation increases during the early acclimation phase and subsequently decreases as elongation dynamics recover. The relatively low coefficient of variation observed at $\mathrm{Gen.\,1}$ may partially reflect sampling bias associated with cell selection at the onset of time-lapse imaging.

To determine whether mechanical confinement alters the distributional form of $L_\mathrm{b}$, we computed empirical cumulative distribution functions (ECDFs) after applying a $z$-score normalization. As shown in Fig.~\ref{fig:homo}(b), the normalized ECDFs collapse onto a common curve across generations, demonstrating that the underlying lognormal distributional shape of $L_\mathrm{b}$ remains largely invariant throughout the acclimation phase. 
Thus, the apparent generational differences in the acclimation phase merely reflect a transient heteroscedasticity, rather than systematic changes in the underlying distributional shape.

The preservation of bounded birth-size statistics with little changes in the mean across generations alongside the previously identified multi-generational growth acclimation necessitates to clarify the role of the underlying size-control mechanism. 
To this end, we established a general framework for generating transient size-control dynamics using numerical simulations of idealized \textit{adder}, \textit{timer}, and \textit{sizer} mechanisms, as a quantitative tool to identify and model a rewired trigger of cell division during mechanical acclimation.

\section{Classification framework for transient size-control dynamics}

\begin{figure*}
    \centering
    \includegraphics[width= .8\textwidth]{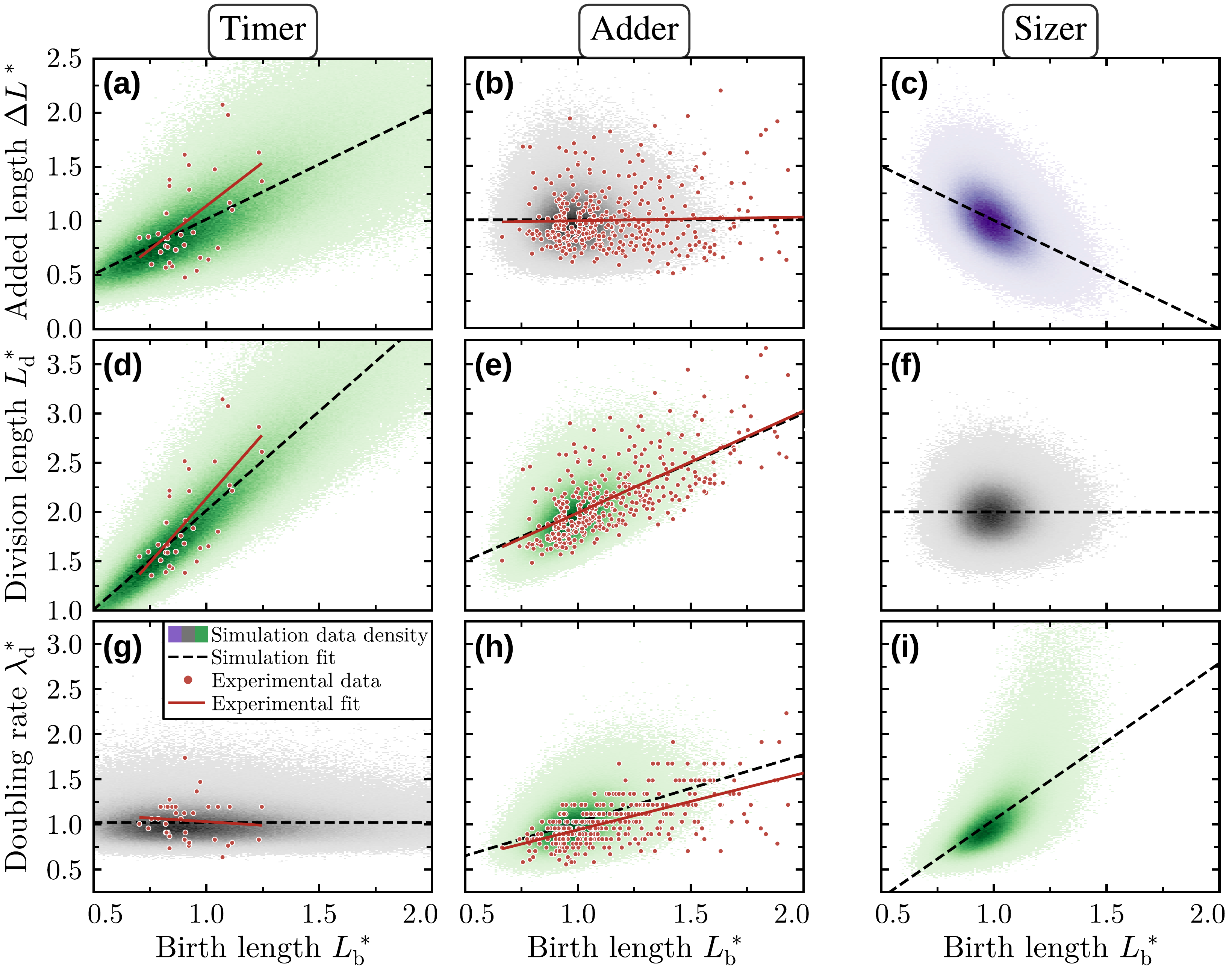}
    \caption{\textbf{Unified classification of canonical size-control signatures.} Two-dimensional density distributions of simulated populations generated under pure \textit{adder} (left column: a, d, g), \textit{sizer} (middle column: b, e, h), and \textit{timer} (right column: c, f, i) control mechanisms. Dashed lines show linear regressions used to extract the slope of each variable pair. Heatmap colors indicate the sign of the correlation: green for positive, purple for negative, and gray for no correlation.
    Experimental data from representative \textit{adder}-like cells (glass--soft gel confinement, $\mathrm{Gen.\,5}$) and \textit{timer}-like cells (glass--soft gel confinement, $\mathrm{Gen.\,1}$) were non-dimensionalized by the mean birth length and doubling rate and overlaid onto the corresponding heatmaps as red dots, together with the linear regression lines (red solid lines).
    The characteristic slopes are stated in Eqs.~\eqref{eq_addersignature} to \eqref{eq_timersignature}.
    }
\label{fig:model_signatures}
\end{figure*}

Different phenomenological strategies have been proposed to characterize bacterial size control, most notably the \textit{sizer}, \textit{timer}, and \textit{adder}, each predicting distinct relationships among birth length $L_\mathrm{b}$, division length $L_\mathrm{d}$, added length $\Delta L$, and division time $\tau_\mathrm{d}$.
In an ideal \textit{sizer}, cells divide upon reaching a characteristic size, such that $L_\mathrm{d}$ is independent of $L_\mathrm{b}$ and $\Delta L$ decreases with increasing $L_\mathrm{b}$. In an ideal \textit{timer}, cells divide after a characteristic time, such that $\tau_\mathrm{d}$ is independent of $L_\mathrm{b}$ and, under exponential elongation, both $L_\mathrm{d}$ and $\Delta L$ increase with $L_\mathrm{b}$. In an ideal \textit{adder}, cells add a characteristic length between birth and division, such that $\Delta L$ is independent of $L_\mathrm{b}$ and $L_\mathrm{d}$ increases linearly with $L_\mathrm{b}$ with a slope of approximately unity (Fig.~\ref{fig:model_signatures}). Under steady growth conditions, single-cell measurements have shown that \textit{adder}-like behavior predominates across a wide range of bacterial species and growth environments \cite{campos2014constant,taheri2015cell}.

These idealized correlation signatures provide a simple framework for distinguishing size-control mechanisms. More general stochastic frameworks describe division through a state-dependent division or hazard rate, providing a unified description from which \textit{sizer}-, \textit{adder}-, or \textit{timer}-like behavior can emerge \cite{osella2014concerted, kennard2016individuality, grilli2017relevant, nieto2020unification}. In experimental populations, however, stochastic variations in growth, division, and cell size can broaden the underlying distributions and blur the characteristic signatures, particularly when sample sizes are limited \cite{facchetti2019reassessment, leTreut2021quantitative, luo2023stochastic, genthon2026cell}. Their interpretation becomes even less straightforward during mechanical acclimation, when the elongation rate itself changes across generations. We therefore constructed a minimal stochastic cell-cycle model implementing pure \textit{sizer}, \textit{timer}, and \textit{adder} control to determine how their characteristic correlation signatures are modified by noise and changing growth rate, and to establish a reference for interpreting the experimentally observed correlations.

In this model, each cell elongates exponentially with a generation-dependent specific growth rate, 
\begin{equation}
\frac{\mathrm{d} l_i}{\mathrm{d}t} = \lambda_i(g) l_i + \sigma l_i \xi_i(t),
\label{eq:minimal_growth}
\end{equation}
where $l_i$ is the length of cell $i$, $g$ labels its generation, $\lambda_i(g)$ is drawn at birth from a generation-dependent distribution, $\sigma$ denotes the noise intensity of single-cell elongation fluctuations, and $\xi_i(t)$ represents uncorrelated Gaussian white noise with zero mean and unit variance. The mean growth rate $\langle \lambda(g) \rangle$ was chosen to follow the experimentally observed recovery after confinement, using the same generation-resolved data shown in Fig.~\ref{fig:ke-shock}, so that the model explicitly incorporates the physiological growth acclimation. Division was then imposed by one of three idealized control rules: an \textit{adder}, in which division occurs after adding a stochastic target length $\Delta L_{\rm target}$; a \textit{sizer}, in which division occurs at a stochastic target division length $L_{\rm target}$; or a \textit{timer}, in which division occurs after a stochastic target interdivision time $\tau_{\rm target}$. At division, the mother cell of length $L_\mathrm{d}$ was partitioned into two daughters with birth lengths $fL_\mathrm{d}$ and $(1-f)L_\mathrm{d}$, where $f$ fluctuates around $1/2$.

To compare model predictions with experimental single-cell data across different growth conditions, we non-dimensionalize the cell-cycle variables using fixed reference scales. 
The characteristic length is defined as $L_0 \equiv \langle \Delta L_{\mathrm{adder}} \rangle$, the mean added length in the \textit{adder} regime ($\mathrm{Gen.\,2 \sim 5}$ for glass--soft gel confinement).The reference doubling time as $\tau_0 \equiv \ln(2)/\lambda_\infty$, where $\lambda_\infty$ is the recovered elongation rate obtained from the exponential acclimation fit. The rescaled variables are
\begin{equation}
    L^* = \frac{L}{L_0}, \qquad
    \tau_\mathrm{d}^* = \frac{\tau_\mathrm{d}}{\tau_0},
    \qquad
    \lambda_\mathrm{d}^* = \lambda_\mathrm{d}\tau_0.
\end{equation}
These reference scales are kept fixed across generations.

The three idealized size-control strategies predict distinct correlation signatures between cell-cycle observables. We characterize these signatures using the slopes 

\begin{align}
    m_{\Delta} &\equiv
    \mathrm{Slope}\,(\Delta L^*,L_\mathrm{b}^*), \\
    m_{\mathrm{d}} &\equiv
    \mathrm{Slope}\,(L_\mathrm{d}^*,L_\mathrm{b}^*), \\
    m_{\lambda} &\equiv
    \mathrm{Slope}\,(\lambda_\mathrm{d}^*,L_\mathrm{b}^*)
\end{align}
expected from the (quasi-) linear relationship between selected pairs of normalized variables.
For an ideal \textit{adder}, the added length $\Delta L$ is independent of birth length $L_\mathrm{b}$, while $L_\mathrm{d}=L_\mathrm{b}+\Delta L$. The corresponding slopes are
\begin{equation}
    m_{\Delta}=0,\qquad
    m_{\mathrm{d}}=1,\qquad
    m_{\lambda}\approx\frac{1}{2}.
    \label{eq_addersignature}
\end{equation}
For an ideal \textit{sizer}, division length $L_\mathrm{d}$ is independent of $L_\mathrm{b}$, while $\Delta L=L_\mathrm{d}-L_\mathrm{b}$. This yields
\begin{equation}
    m_{\Delta}=-1,\qquad
    m_{\mathrm{d}}=0,\qquad
    m_{\lambda}\approx 1.
    \label{eq_sizersignature}
\end{equation}
For an ideal \textit{timer}, division time $\tau_\mathrm{d}$ is independent of $L_\mathrm{b}$ and, under exponential elongation, $L_\mathrm{d}=L_\mathrm{b} \exp(\lambda_\mathrm{e}\tau_\mathrm{d})$, therefore
\begin{equation}
    m_{\Delta}\approx 1,\qquad
    m_{\mathrm{d}}\approx 2,\qquad
    m_{\lambda}=0.
    \label{eq_timersignature}
\end{equation}

Note that the predicted values of $m_{\lambda}$ for the \textit{adder} and \textit{sizer} correspond to local linear approximations near the reference birth length, as the division-rate relationships are generally nonlinear. Further details can be found in the Supplementary Material.

The idealized slopes in Eqs.~\eqref{eq_addersignature} to \eqref{eq_timersignature} provide a compact classification scheme for distinguishing between the \textit{adder}, \textit{sizer}, and \textit{timer} size-control mechanisms.
They serve as benchmarks for interpreting the experimental correlation signatures by fitting linear regression models (see Fig.~\ref{fig:model_signatures}), rather than as an exact description of every generation. Detailed derivations and statistical comparisons are provided in the Supplementary Material.

Our simulation framework allows the physiological recovery of growth rate and the structural rule of division control to be varied independently.
We thus further examined whether correlations between elongation rate $\lambda_\mathrm{e}$ and birth length $L_\mathrm{b}$ could confound the classification of size-control dynamics. Numerical simulations show that \textit{adder}- and \textit{sizer}-like regulation remain robustly distinguishable over a broad range of coupling strengths, whereas \textit{timer}-like signatures can mimic \textit{adder}-like slopes but only under unrealistically strong negative $\lambda_\mathrm{e}$--$L_\mathrm{b}$ correlations (Supplementary Fig.~S18). These results demonstrate that the proposed classification framework is robust to the level of growth-rate heterogeneity observed experimentally.

\section{Transient departure from canonical \textit{adder} regulation}

\begin{figure*}
	\centering
	\includegraphics[width=\textwidth]{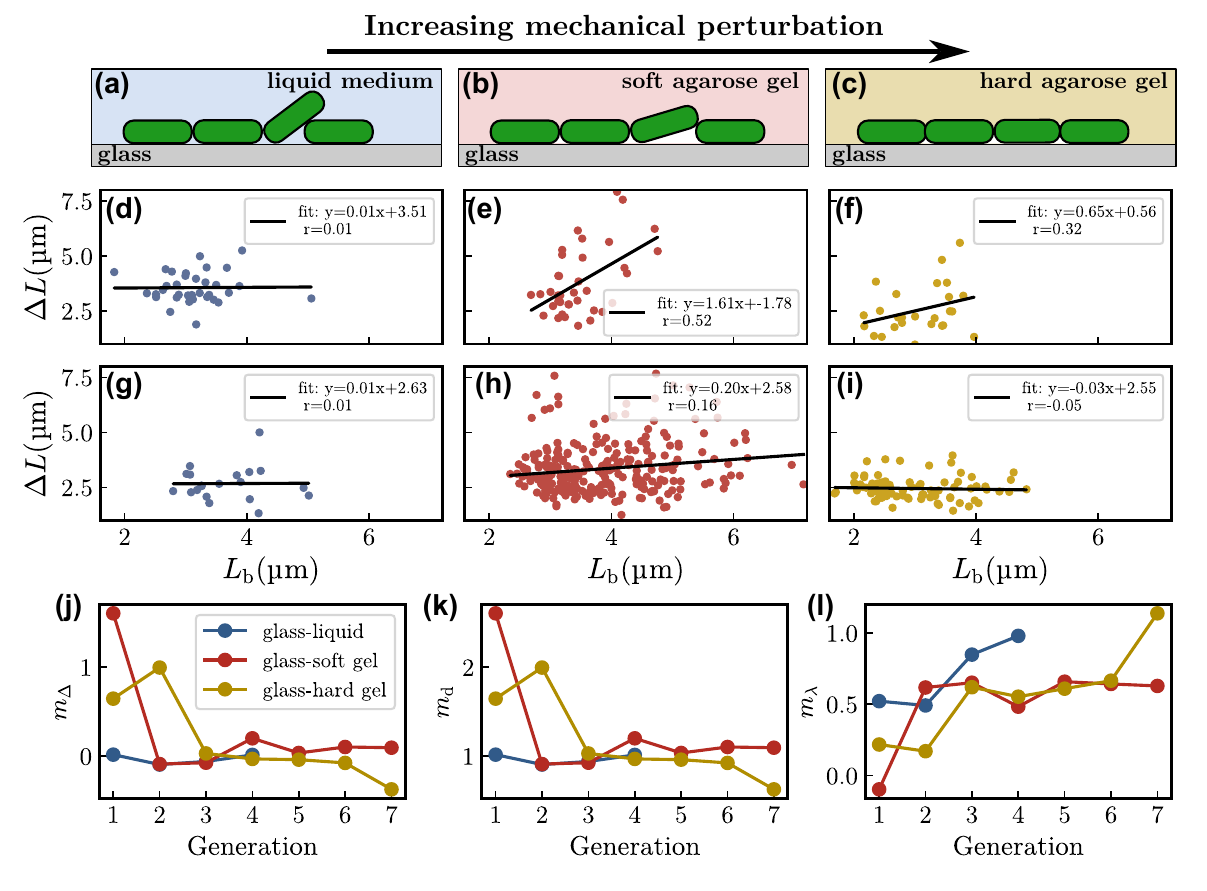} \\
	\caption{
\textbf{Mechanical-stress-dependent rewiring of size-control signatures.} \textbf{(a)--(c)} Experimental conditions of increasing confinement strength: glass--LB, glass--1.5\% agarose gel, and glass--6\% agarose gel. \textbf{(d)--(f)} Correlations between $\Delta L$ and $L_\mathrm{b}$ during the early acclimation phase ($\mathrm{Gen.\,1}$). \textbf{(g)--(i)} Corresponding correlations after recovery ($\mathrm{Gen.\,4}$). \textbf{(j)--(l)} Generation-resolved evolution of $m_\Delta$, $m_\mathrm{d}$, and $m_\lambda$. Note the rescaled variables $\lambda_\mathrm{d}^* = \lambda_\mathrm{d} \cdot \tau_0$, $L_\mathrm{b}^* = L_\mathrm{b}/\langle \Delta L \rangle$. Intermediate and strong confinement transiently shift cells away from canonical \textit{adder} regulation before recovery at later generations. See Supplementary Figs.~S10--S13 for the full generation-resolved correlation signatures.
}
	\label{fig:timer} 
\end{figure*}


Applying the classification framework to experimental data reveals a confinement-strength-dependent remodeling of bacterial size-control dynamics (Fig.~\ref{fig:timer}). 
The canonical \textit{adder} behavior, with $\Delta L$ largely independent of $L_\mathrm{b}$ and the $L_\mathrm{d}$--$L_\mathrm{b}$ slope remaining near unity, is observed or under all confinement conditions apart from an initial transient under gel confinement (Fig.~\ref{fig:timer}(j--l). 
Under weak confinement (glass--liquid medium), cell-size statistics show these typical \textit{adder} signatures throughout the experiment (Fig.~\ref{fig:timer}(d,g), Supplementary Fig.~S10).
Consistently, statistical tests of the generation-resolved regression slopes show no significant departure from the ideal \textit{adder} prediction for added length, while rejecting the \textit{timer} prediction across all measured generations. 

By contrast, glass--soft gel ($1.5\%$ agarose) confinement induces a clear transient \textit{timer}-like shift during $\mathrm{Gen.\,1}$: $\Delta L$ becomes positively correlated with $L_\mathrm{b}$, the $L_\mathrm{d}$--$L_\mathrm{b}$ slope exceeds unity, and the correlation between $\tau_\mathrm{d}$ and $L_\mathrm{b}$ is weakened compared to all later generations (Fig.~\ref{fig:timer}(e, j--l), Supplementary Figs.~S11--S12). 
This transient shift is statistically supported by the added-length slope ($m_{\Delta}=1.61$, $95\%$ CI $[0.74,2.47]$), which significantly departs from the \textit{adder} prediction ($m_{\Delta}=0$, $p<0.001$) while remaining compatible with the ideal \textit{timer} prediction ($m_{\Delta}=1$). The rescaled division-rate slope ($m_{\lambda}=-0.10$) provides complementary evidence, significantly departing from the \textit{adder} reference value ($m_{\lambda} \approx 0.5$) while remaining statistically compatible with the zero value of a \textit{timer}.
These signatures indicate a transient departure from canonical \textit{adder} regulation, consistent with partial decoupling between growth and division. 
Previous studies have shown that bacterial size-control behavior can vary with physiological state: \textit{adder}-like regulation is commonly observed during faster growth, whereas slower growth can produce more \textit{sizer}-like behavior \cite{wallden2016synchronization, lee2019microfluidic, nieto2024mechanisms}. In our experiments, however, the transient departure from \textit{adder} regulation is \textit{timer}-like rather than \textit{sizer}-like.

Under glass--hard gel ($6\%$ agarose) confinement, a transient shift toward \textit{timer}-like regulation is observed in both the first and second generations (Fig.~\ref{fig:timer}(f, j--l), Supplementary Fig.~S13). 
Both generations exhibit positive added-length slopes ($m_{\Delta}=0.65$ and $1.00$, respectively), together with rescaled division-rate slopes ($m_{\lambda}=0.22$ and $0.17$, respectively) approaching the ideal \textit{timer} prediction ($m_{\lambda}=0$). 
Although the limited sample size ($n=26$) and substantial variability in $\mathrm{Gen.\,1}$ result in wide confidence intervals that prevent a statistically significant distinction between the two ideal mechanisms, $\mathrm{Gen.\,2}$ shows significant departures from the \textit{adder} hypothesis for both observables ($p<10^{-4}$ for $m_{\Delta}$ and $p=0.002$ for $m_{\lambda}$), while remaining statistically consistent with the \textit{timer} hypothesis. 
These results support a \textit{timer}-like response extending over two generations. 
This prolonged transient and the increased heterogeneity may reflect stronger mechanical confinement and its effects on elongation and division dynamics. Moreover, the smaller cell size observed under hard-gel confinement (Supplementary Fig.~S5) is consistent with the growth slowdown and size reduction previously reported in crowded bacterial colonies \cite{wittmann2023collective}.

Across all three confinement conditions, \textit{adder}-like size regulation is observed or recovered at different generations depending on confinement strength: from $\mathrm{Gen.\,1}$ under glass--liquid conditions, $\mathrm{Gen.\,2}$ under glass--soft gel confinement, and $\mathrm{Gen.\,3}$ under glass--hard gel confinement. In these regimes, the $\Delta L$--$L_\mathrm{b}$ slope approaches zero (Fig.~\ref{fig:timer}(g--j)), while the $L_\mathrm{d}$--$L_\mathrm{b}$ slope approaches one (Fig.~\ref{fig:timer}(k)). A modest negative added-length slope in $\mathrm{Gen.,7}$ under hard-gel confinement suggests a possible late-stage \textit{sizer}-like tendency. 
Overall, these results indicate that stronger confinement prolongs the transient departure from \textit{adder}-like regulation, followed by its gradual recovery. 
Notably, this recovery precedes the full recovery of the elongation rate (Fig.~\ref{fig:timer}(j--l)), indicating that size-control coordination and bulk growth physiology relax on distinct timescales. Such a separation is consistent with nutrient-shift studies showing that division-related parameters can adjust on a timescale distinct from that of the overall growth rate \cite{panlilio2021threshold, shi2021precise,buke2022ppgpp, pahujani2026adaptive}.

\begin{figure*}
    \centering
    \includegraphics[width=\textwidth]{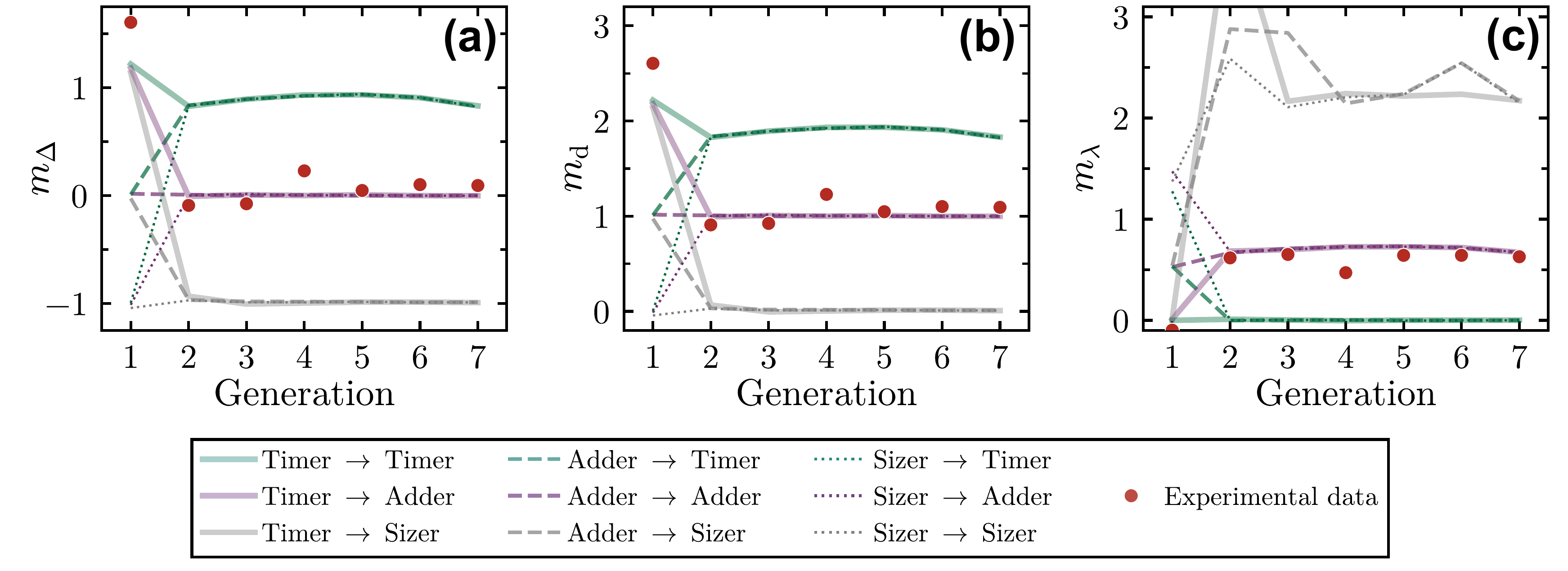}
    \caption{\textbf{A \textit{timer}-like-to-\textit{adder}-like transformation reproduces the experimental transient correlation dynamics.} Comparison between experimental correlation slopes and simulations with candidate static and switching control models. 
    Allowing the growth rate to recover over generations is insufficient to reproduce the measured slope trajectories when the control rule is fixed. Among the tested two-state transitions, only the \textit{timer}-like-to-\textit{adder}-like transformation captures the experimentally observed shift from early \textit{timer}-like correlations to late \textit{adder}-like regulation.
    }
    \label{fig:model_transition}
\end{figure*}

We further considered whether these transient correlation dynamics could simply arise from the gradual recovery of the elongation rate, rather than reflecting a genuine departure from the underlying \textit{adder} size-control mechanism. To test this possibility, we simulated populations in which the experimentally measured generation-dependent elongation-rate statistics were imposed, but the division rule was fixed as an \textit{adder}, \textit{sizer}, or \textit{timer}. Although these models reproduced the changing growth and division timescales, they failed to predict the observed evolution of the experimental correlation signatures (Fig.~\ref{fig:model_transition}), 
demonstrating that growth-rate recovery alone cannot account for the experimental observations. 

We also extended our models by allowing the effective size-control rule to change dynamically. 
Among all two-state transitions between \textit{adder}-, \textit{sizer}-, and \textit{timer}-like regulation, only a transient \textit{timer}-like-to-\textit{adder}-like transition reproduced the experimental evolution of all three classification observables (Fig.~\ref{fig:model_transition}, purple solid line). The model captures both the early \textit{timer}-like signatures, characterized by positive $\Delta L$--$L_\mathrm{b}$ correlations and super-unity $L_\mathrm{d}$--$L_\mathrm{b}$ slopes, and the subsequent restoration of canonical \textit{adder} behavior. These results indicate that mechanical confinement does not merely rescale bacterial growth dynamics, but transiently remodels the effective growth--division coupling before canonical \textit{adder} regulation is re-established.

Finally, we demonstrated that the colony can maintain the population-level size homeostasis despite the transient \textit{timer}-like behavior inferred from the correlation analysis. To this end, we simulated a generalized \textit{timer}-to-\textit{adder} transition occurring beyond the first generation. Although the transient \textit{timer} state results in a temporary broadening of the birth-size distribution, it remains bounded and rapidly converges back to a universal steady-state after the \textit{adder} behavior is restored (Supplementary Figs.~S16--S17). Therefore, a transient \textit{timer}-like regulation persisting over only a few generations does not destabilize the global birth-size statistics, establishing this dynamic size-control mechanism as an effective strategy for mechanical acclimation to new environments.

\section{Discussion}

Bacteria frequently inhabit mechanically structured soft environments in which growth occurs under surface contact, crowding, and confinement. We have found that entry into such environments initiates a multi-generational reorganization of growth and division dynamics, highlighting mechanical acclimation as an important aspect of bacterial physiology beyond steady-state conditions.
We use controlled surface attachment and agarose confinement to isolate this transition and follow its consequences across generations. 

Our results show that transferring exponentially growing \textit{E. coli} from liquid culture to surface-associated or confined environments induces an immediate reduction in elongation rate, followed by multi-generational acclimation of growth and division dynamics. Even glass attachment in liquid medium produces a measurable response, which becomes stronger and more prolonged with increasing stiffness of contacting gels. These findings demonstrate that surface attachment and confinement are not merely passive imaging conditions but physical perturbations that trigger dynamic physiological responses.

Under gel confinements, mechanical acclimation is also accompanied by a transient shift from canonical \textit{adder}-like regulation toward \textit{timer}-like correlations, followed by the restoration of \textit{adder}-like behavior. Under soft-gel confinement, the shift was detected in $\mathrm{Gen.\,1}$, whereas under hard-gel confinement, the slope estimates suggested a two-generation transient. 

Despite this transient reorganization, birth-size distributions remain bounded throughout acclimation. Simulations further show that \textit{timer}-like regulation lasting only a few generations does not produce the size divergence expected under a sustained \textit{timer} mechanism. Thus, transient changes in growth--division coordination can occur without disrupting size homeostasis.

Notably, \textit{adder}-like regulation is restored before the elongation rate fully recovers, indicating distinct recovery timescales for size-control coordination and cellular growth. Our correlation-based framework provides a general approach for characterizing such transient responses under non-steady growth conditions. More broadly, our results highlight bacterial size control as a dynamical process beyond its conventional steady-state description: following an environmental change, growth--division correlations can be transiently reorganized and subsequently restored on timescales distinct from those governing the recovery of cellular growth.

\begin{acknowledgments}
We acknowledge financial support from the Institute for Advanced Studies, University of Luxembourg (AUDACITY Grant IAS-21/CAMEOS to A.S.), the Marie Sk\l{}odowska-Curie Actions Individual Fellowship (Project 101110587-MINIMA to C.J.), and the Luxembourg National Research Fund (FNR) through ATTRACT Investigator Grant (A17/MS/ 11572821/MBRACE) and a CORE Grant (C19/MS/13719464/TOPOFLUME/Sengupta) to A.S. A.S. thanks Nicola Storelli and Francesco Origgi for insightful discussions on bacterial physiology. G.H.P.N., R.W., and H.L. acknowledge funding from the Deutsche Forschungsgemeinschaft (DFG, German Research Foundation) through the Priority Programme SPP 2265 (Grant Nos. WI 5527/1-2, and LO 418/25-2). 

\textbf{AI Usage Disclosure:} OpenAI ChatGPT (GPT-5.6 Sol) was used during manuscript preparation to assist with translation, language editing, condensation, organization of the text, and refinement of the presentation and logical structure of the arguments. All AI-assisted content was critically reviewed, verified, and revised by the authors, who take full responsibility for the manuscript. Generative AI was not used for experimental data analysis or figure generation.

\end{acknowledgments}

\section*{Data Availability Statement}
The data supporting the findings of this study, including the numerical data underlying the figures, have been deposited in Zenodo \cite{dataset}. During peer review, the data are available to the editors and reviewers through a private repository link. The dataset will be made publicly available through Zenodo upon publication of this article.

\bibliography{mechno-ref-abbrev-prx}

\begin{thebibliography}{67}%
\makeatletter
\providecommand \@ifxundefined [1]{%
 \@ifx{#1\undefined}
}%
\providecommand \@ifnum [1]{%
 \ifnum #1\expandafter \@firstoftwo
 \else \expandafter \@secondoftwo
 \fi
}%
\providecommand \@ifx [1]{%
 \ifx #1\expandafter \@firstoftwo
 \else \expandafter \@secondoftwo
 \fi
}%
\providecommand \natexlab [1]{#1}%
\providecommand \enquote  [1]{``#1''}%
\providecommand \bibnamefont  [1]{#1}%
\providecommand \bibfnamefont [1]{#1}%
\providecommand \citenamefont [1]{#1}%
\providecommand \href@noop [0]{\@secondoftwo}%
\providecommand \href [0]{\begingroup \@sanitize@url \@href}%
\providecommand \@href[1]{\@@startlink{#1}\@@href}%
\providecommand \@@href[1]{\endgroup#1\@@endlink}%
\providecommand \@sanitize@url [0]{\catcode `\\12\catcode `\$12\catcode `\&12\catcode `\#12\catcode `\^12\catcode `\_12\catcode `\%12\relax}%
\providecommand \@@startlink[1]{}%
\providecommand \@@endlink[0]{}%
\providecommand \url  [0]{\begingroup\@sanitize@url \@url }%
\providecommand \@url [1]{\endgroup\@href {#1}{\urlprefix }}%
\providecommand \urlprefix  [0]{URL }%
\providecommand \Eprint [0]{\href }%
\providecommand \doibase [0]{https://doi.org/}%
\providecommand \selectlanguage [0]{\@gobble}%
\providecommand \bibinfo  [0]{\@secondoftwo}%
\providecommand \bibfield  [0]{\@secondoftwo}%
\providecommand \translation [1]{[#1]}%
\providecommand \BibitemOpen [0]{}%
\providecommand \bibitemStop [0]{}%
\providecommand \bibitemNoStop [0]{.\EOS\space}%
\providecommand \EOS [0]{\spacefactor3000\relax}%
\providecommand \BibitemShut  [1]{\csname bibitem#1\endcsname}%
\let\auto@bib@innerbib\@empty
\bibitem [{\citenamefont {Flemming}\ and\ \citenamefont {Wuertz}(2019)}]{flemming2019bacteria}%
  \BibitemOpen
  \bibfield  {author} {\bibinfo {author} {\bibfnamefont {H.-C.}\ \bibnamefont {Flemming}}\ and\ \bibinfo {author} {\bibfnamefont {S.}~\bibnamefont {Wuertz}},\ }\href {https://doi.org/10.1038/s41579-019-0158-9} {\bibfield  {journal} {\bibinfo  {journal} {Nat. Rev. Microbiol.}\ }\textbf {\bibinfo {volume} {17}},\ \bibinfo {pages} {247} (\bibinfo {year} {2019})}\BibitemShut {NoStop}%
\bibitem [{\citenamefont {Bar-On}\ \emph {et~al.}(2018)\citenamefont {Bar-On}, \citenamefont {Phillips},\ and\ \citenamefont {Milo}}]{bar2018biomass}%
  \BibitemOpen
  \bibfield  {author} {\bibinfo {author} {\bibfnamefont {Y.~M.}\ \bibnamefont {Bar-On}}, \bibinfo {author} {\bibfnamefont {R.}~\bibnamefont {Phillips}},\ and\ \bibinfo {author} {\bibfnamefont {R.}~\bibnamefont {Milo}},\ }\href {https://doi.org/10.1073/pnas.1711842115} {\bibfield  {journal} {\bibinfo  {journal} {Proc. Natl. Acad. Sci. USA}\ }\textbf {\bibinfo {volume} {115}},\ \bibinfo {pages} {6506} (\bibinfo {year} {2018})}\BibitemShut {NoStop}%
\bibitem [{\citenamefont {Jin}\ and\ \citenamefont {Sengupta}(2024)}]{jin2024microbes}%
  \BibitemOpen
  \bibfield  {author} {\bibinfo {author} {\bibfnamefont {C.}~\bibnamefont {Jin}}\ and\ \bibinfo {author} {\bibfnamefont {A.}~\bibnamefont {Sengupta}},\ }\href {https://doi.org/10.1007/s12551-024-01185-7} {\bibfield  {journal} {\bibinfo  {journal} {Biophys. Rev.}\ }\textbf {\bibinfo {volume} {16}},\ \bibinfo {pages} {173} (\bibinfo {year} {2024})}\BibitemShut {NoStop}%
\bibitem [{\citenamefont {Ebrahimi}\ and\ \citenamefont {Or}(2014)}]{ebrahimi2014}%
  \BibitemOpen
  \bibfield  {author} {\bibinfo {author} {\bibfnamefont {A.}~\bibnamefont {Ebrahimi}}\ and\ \bibinfo {author} {\bibfnamefont {D.}~\bibnamefont {Or}},\ }\href {https://doi.org/10.1002/2014WR015897} {\bibfield  {journal} {\bibinfo  {journal} {Water Resour. Res.}\ }\textbf {\bibinfo {volume} {50}},\ \bibinfo {pages} {7406} (\bibinfo {year} {2014})}\BibitemShut {NoStop}%
\bibitem [{\citenamefont {Kondakova}\ \emph {et~al.}(2013)\citenamefont {Kondakova}, \citenamefont {Verkhovtseva},\ and\ \citenamefont {Ostroumov}}]{kondakova2013study}%
  \BibitemOpen
  \bibfield  {author} {\bibinfo {author} {\bibfnamefont {G.}~\bibnamefont {Kondakova}}, \bibinfo {author} {\bibfnamefont {N.}~\bibnamefont {Verkhovtseva}},\ and\ \bibinfo {author} {\bibfnamefont {S.}~\bibnamefont {Ostroumov}},\ }\href {https://doi.org/10.3103/S0096392513030061} {\bibfield  {journal} {\bibinfo  {journal} {Mosc. Univ. Biol. Sci. Bull.}\ }\textbf {\bibinfo {volume} {68}},\ \bibinfo {pages} {119} (\bibinfo {year} {2013})}\BibitemShut {NoStop}%
\bibitem [{\citenamefont {Philippot}\ \emph {et~al.}(2024)\citenamefont {Philippot}, \citenamefont {Chenu}, \citenamefont {Kappler}, \citenamefont {Rillig},\ and\ \citenamefont {Fierer}}]{philippot2023}%
  \BibitemOpen
  \bibfield  {author} {\bibinfo {author} {\bibfnamefont {L.}~\bibnamefont {Philippot}}, \bibinfo {author} {\bibfnamefont {C.}~\bibnamefont {Chenu}}, \bibinfo {author} {\bibfnamefont {A.}~\bibnamefont {Kappler}}, \bibinfo {author} {\bibfnamefont {M.~C.}\ \bibnamefont {Rillig}},\ and\ \bibinfo {author} {\bibfnamefont {N.}~\bibnamefont {Fierer}},\ }\href {https://doi.org/10.1038/s41579-023-00980-5} {\bibfield  {journal} {\bibinfo  {journal} {Nat. Rev. Microbiol.}\ }\textbf {\bibinfo {volume} {22}},\ \bibinfo {pages} {226} (\bibinfo {year} {2024})}\BibitemShut {NoStop}%
\bibitem [{\citenamefont {Heinemann}\ \emph {et~al.}(2021)\citenamefont {Heinemann}, \citenamefont {Alcalde}, \citenamefont {Miocic}, \citenamefont {Hangx}, \citenamefont {Kallmeyer}, \citenamefont {Ostertag-Henning}, \citenamefont {Hassanpouryouzband}, \citenamefont {Thaysen}, \citenamefont {Strobel}, \citenamefont {Schmidt-Hattenberger} \emph {et~al.}}]{heinemann2021}%
  \BibitemOpen
  \bibfield  {author} {\bibinfo {author} {\bibfnamefont {N.}~\bibnamefont {Heinemann}}, \bibinfo {author} {\bibfnamefont {J.}~\bibnamefont {Alcalde}}, \bibinfo {author} {\bibfnamefont {J.~M.}\ \bibnamefont {Miocic}}, \bibinfo {author} {\bibfnamefont {S.~J.~T.}\ \bibnamefont {Hangx}}, \bibinfo {author} {\bibfnamefont {J.}~\bibnamefont {Kallmeyer}}, \bibinfo {author} {\bibfnamefont {C.}~\bibnamefont {Ostertag-Henning}}, \bibinfo {author} {\bibfnamefont {A.}~\bibnamefont {Hassanpouryouzband}}, \bibinfo {author} {\bibfnamefont {E.~M.}\ \bibnamefont {Thaysen}}, \bibinfo {author} {\bibfnamefont {G.~J.}\ \bibnamefont {Strobel}}, \bibinfo {author} {\bibfnamefont {C.}~\bibnamefont {Schmidt-Hattenberger}}, \emph {et~al.},\ }\href {https://doi.org/10.1039/D0EE03536J} {\bibfield  {journal} {\bibinfo  {journal} {Energy Environ. Sci.}\ }\textbf {\bibinfo {volume} {14}},\ \bibinfo {pages} {853} (\bibinfo {year} {2021})}\BibitemShut {NoStop}%
\bibitem [{\citenamefont {Muhammed}\ \emph {et~al.}(2022)\citenamefont {Muhammed}, \citenamefont {Haq}, \citenamefont {Al~Shehri}, \citenamefont {Al-Ahmed}, \citenamefont {Rahman},\ and\ \citenamefont {Zaman}}]{muhammed2022}%
  \BibitemOpen
  \bibfield  {author} {\bibinfo {author} {\bibfnamefont {N.~S.}\ \bibnamefont {Muhammed}}, \bibinfo {author} {\bibfnamefont {B.}~\bibnamefont {Haq}}, \bibinfo {author} {\bibfnamefont {D.}~\bibnamefont {Al~Shehri}}, \bibinfo {author} {\bibfnamefont {A.}~\bibnamefont {Al-Ahmed}}, \bibinfo {author} {\bibfnamefont {M.~M.}\ \bibnamefont {Rahman}},\ and\ \bibinfo {author} {\bibfnamefont {E.}~\bibnamefont {Zaman}},\ }\href {https://doi.org/10.1016/j.egyr.2021.12.002} {\bibfield  {journal} {\bibinfo  {journal} {Energy Rep.}\ }\textbf {\bibinfo {volume} {8}},\ \bibinfo {pages} {461} (\bibinfo {year} {2022})}\BibitemShut {NoStop}%
\bibitem [{\citenamefont {Kirch}\ \emph {et~al.}(2012)\citenamefont {Kirch}, \citenamefont {Schneider}, \citenamefont {Abou}, \citenamefont {Hopf}, \citenamefont {Schaefer}, \citenamefont {Schneider}, \citenamefont {Schall}, \citenamefont {Wagner},\ and\ \citenamefont {Lehr}}]{kirch2012}%
  \BibitemOpen
  \bibfield  {author} {\bibinfo {author} {\bibfnamefont {J.}~\bibnamefont {Kirch}}, \bibinfo {author} {\bibfnamefont {A.}~\bibnamefont {Schneider}}, \bibinfo {author} {\bibfnamefont {B.}~\bibnamefont {Abou}}, \bibinfo {author} {\bibfnamefont {A.}~\bibnamefont {Hopf}}, \bibinfo {author} {\bibfnamefont {U.}~\bibnamefont {Schaefer}}, \bibinfo {author} {\bibfnamefont {M.}~\bibnamefont {Schneider}}, \bibinfo {author} {\bibfnamefont {C.}~\bibnamefont {Schall}}, \bibinfo {author} {\bibfnamefont {C.}~\bibnamefont {Wagner}},\ and\ \bibinfo {author} {\bibfnamefont {C.}~\bibnamefont {Lehr}},\ }\href {https://doi.org/10.1073/pnas.1214066109} {\bibfield  {journal} {\bibinfo  {journal} {Proc. Natl. Acad. Sci. USA}\ }\textbf {\bibinfo {volume} {109}},\ \bibinfo {pages} {18355} (\bibinfo {year} {2012})}\BibitemShut {NoStop}%
\bibitem [{\citenamefont {Wheeler}\ \emph {et~al.}(2019)\citenamefont {Wheeler}, \citenamefont {Cárcamo-Oyarce}, \citenamefont {Turner}, \citenamefont {Dellos-Nolan}, \citenamefont {Co}, \citenamefont {Lehoux}, \citenamefont {Cummings}, \citenamefont {Wozniak},\ and\ \citenamefont {Ribbeck}}]{wheeler2019}%
  \BibitemOpen
  \bibfield  {author} {\bibinfo {author} {\bibfnamefont {K.}~\bibnamefont {Wheeler}}, \bibinfo {author} {\bibfnamefont {G.}~\bibnamefont {Cárcamo-Oyarce}}, \bibinfo {author} {\bibfnamefont {B.}~\bibnamefont {Turner}}, \bibinfo {author} {\bibfnamefont {S.}~\bibnamefont {Dellos-Nolan}}, \bibinfo {author} {\bibfnamefont {J.}~\bibnamefont {Co}}, \bibinfo {author} {\bibfnamefont {S.}~\bibnamefont {Lehoux}}, \bibinfo {author} {\bibfnamefont {R.}~\bibnamefont {Cummings}}, \bibinfo {author} {\bibfnamefont {D.}~\bibnamefont {Wozniak}},\ and\ \bibinfo {author} {\bibfnamefont {K.}~\bibnamefont {Ribbeck}},\ }\href {https://doi.org/10.1038/s41564-019-0581-8} {\bibfield  {journal} {\bibinfo  {journal} {Nat. Microbiol.}\ }\textbf {\bibinfo {volume} {4}},\ \bibinfo {pages} {2146} (\bibinfo {year} {2019})}\BibitemShut {NoStop}%
\bibitem [{\citenamefont {Bansil}\ and\ \citenamefont {Turner}(2018)}]{bansil2018}%
  \BibitemOpen
  \bibfield  {author} {\bibinfo {author} {\bibfnamefont {R.}~\bibnamefont {Bansil}}\ and\ \bibinfo {author} {\bibfnamefont {B.}~\bibnamefont {Turner}},\ }\href {https://doi.org/10.1016/j.addr.2017.09.023} {\bibfield  {journal} {\bibinfo  {journal} {Adv. Drug Deliv. Rev.}\ }\textbf {\bibinfo {volume} {124}},\ \bibinfo {pages} {3} (\bibinfo {year} {2018})}\BibitemShut {NoStop}%
\bibitem [{\citenamefont {Wu}\ \emph {et~al.}(2023)\citenamefont {Wu}, \citenamefont {Wheeler}, \citenamefont {C{\'a}rcamo-Oyarce}, \citenamefont {Aoki}, \citenamefont {McShane}, \citenamefont {Datta}, \citenamefont {Mark~Welch}, \citenamefont {Tiemeyer}, \citenamefont {Griffen},\ and\ \citenamefont {Ribbeck}}]{wu2023mucin}%
  \BibitemOpen
  \bibfield  {author} {\bibinfo {author} {\bibfnamefont {C.~M.}\ \bibnamefont {Wu}}, \bibinfo {author} {\bibfnamefont {K.~M.}\ \bibnamefont {Wheeler}}, \bibinfo {author} {\bibfnamefont {G.}~\bibnamefont {C{\'a}rcamo-Oyarce}}, \bibinfo {author} {\bibfnamefont {K.}~\bibnamefont {Aoki}}, \bibinfo {author} {\bibfnamefont {A.}~\bibnamefont {McShane}}, \bibinfo {author} {\bibfnamefont {S.~S.}\ \bibnamefont {Datta}}, \bibinfo {author} {\bibfnamefont {J.~L.}\ \bibnamefont {Mark~Welch}}, \bibinfo {author} {\bibfnamefont {M.}~\bibnamefont {Tiemeyer}}, \bibinfo {author} {\bibfnamefont {A.~L.}\ \bibnamefont {Griffen}},\ and\ \bibinfo {author} {\bibfnamefont {K.}~\bibnamefont {Ribbeck}},\ }\href {https://doi.org/10.1038/s41522-023-00378-4} {\bibfield  {journal} {\bibinfo  {journal} {npj Biofilms Microbiomes}\ }\textbf {\bibinfo {volume} {9}},\ \bibinfo {pages} {11} (\bibinfo {year} {2023})}\BibitemShut {NoStop}%
\bibitem [{\citenamefont {Dang}\ and\ \citenamefont {Lovell}(2016)}]{dang2016}%
  \BibitemOpen
  \bibfield  {author} {\bibinfo {author} {\bibfnamefont {H.}~\bibnamefont {Dang}}\ and\ \bibinfo {author} {\bibfnamefont {C.~R.}\ \bibnamefont {Lovell}},\ }\href {https://doi.org/10.1128/MMBR.00037-15} {\bibfield  {journal} {\bibinfo  {journal} {Microbiol. Mol. Biol. Rev.}\ }\textbf {\bibinfo {volume} {80}},\ \bibinfo {pages} {91} (\bibinfo {year} {2016})}\BibitemShut {NoStop}%
\bibitem [{\citenamefont {Alcolombri}\ \emph {et~al.}(2021)\citenamefont {Alcolombri}, \citenamefont {Peaudecerf}, \citenamefont {Fernandez}, \citenamefont {Behrendt}, \citenamefont {Lee},\ and\ \citenamefont {Stocker}}]{alcolombri2021sinking}%
  \BibitemOpen
  \bibfield  {author} {\bibinfo {author} {\bibfnamefont {U.}~\bibnamefont {Alcolombri}}, \bibinfo {author} {\bibfnamefont {F.~J.}\ \bibnamefont {Peaudecerf}}, \bibinfo {author} {\bibfnamefont {V.~I.}\ \bibnamefont {Fernandez}}, \bibinfo {author} {\bibfnamefont {L.}~\bibnamefont {Behrendt}}, \bibinfo {author} {\bibfnamefont {K.~S.}\ \bibnamefont {Lee}},\ and\ \bibinfo {author} {\bibfnamefont {R.}~\bibnamefont {Stocker}},\ }\href {https://doi.org/10.1038/s41561-021-00817-x} {\bibfield  {journal} {\bibinfo  {journal} {Nat. Geosci.}\ }\textbf {\bibinfo {volume} {14}},\ \bibinfo {pages} {775} (\bibinfo {year} {2021})}\BibitemShut {NoStop}%
\bibitem [{\citenamefont {Borer}\ \emph {et~al.}(2023)\citenamefont {Borer}, \citenamefont {Zhang}, \citenamefont {Baker}, \citenamefont {O'Toole},\ and\ \citenamefont {Babbin}}]{borer2023}%
  \BibitemOpen
  \bibfield  {author} {\bibinfo {author} {\bibfnamefont {B.}~\bibnamefont {Borer}}, \bibinfo {author} {\bibfnamefont {I.}~\bibnamefont {Zhang}}, \bibinfo {author} {\bibfnamefont {A.}~\bibnamefont {Baker}}, \bibinfo {author} {\bibfnamefont {G.}~\bibnamefont {O'Toole}},\ and\ \bibinfo {author} {\bibfnamefont {A.}~\bibnamefont {Babbin}},\ }\href {https://doi.org/10.1093/pnasnexus/pgac311} {\bibfield  {journal} {\bibinfo  {journal} {PNAS Nexus}\ }\textbf {\bibinfo {volume} {2}},\ \bibinfo {pages} {pgac311} (\bibinfo {year} {2023})}\BibitemShut {NoStop}%
\bibitem [{\citenamefont {Alcolombri}\ \emph {et~al.}(2025)\citenamefont {Alcolombri}, \citenamefont {Nissan}, \citenamefont {S{\l}omka}, \citenamefont {Charlton}, \citenamefont {Secchi}, \citenamefont {Short}, \citenamefont {Lee}, \citenamefont {Peaudecerf}, \citenamefont {Baumgartner}, \citenamefont {Sichert} \emph {et~al.}}]{alcolombri2025biogel}%
  \BibitemOpen
  \bibfield  {author} {\bibinfo {author} {\bibfnamefont {U.}~\bibnamefont {Alcolombri}}, \bibinfo {author} {\bibfnamefont {A.}~\bibnamefont {Nissan}}, \bibinfo {author} {\bibfnamefont {J.}~\bibnamefont {S{\l}omka}}, \bibinfo {author} {\bibfnamefont {S.}~\bibnamefont {Charlton}}, \bibinfo {author} {\bibfnamefont {E.}~\bibnamefont {Secchi}}, \bibinfo {author} {\bibfnamefont {I.}~\bibnamefont {Short}}, \bibinfo {author} {\bibfnamefont {K.~S.}\ \bibnamefont {Lee}}, \bibinfo {author} {\bibfnamefont {F.~J.}\ \bibnamefont {Peaudecerf}}, \bibinfo {author} {\bibfnamefont {D.~A.}\ \bibnamefont {Baumgartner}}, \bibinfo {author} {\bibfnamefont {A.}~\bibnamefont {Sichert}}, \emph {et~al.},\ }\href {https://doi.org/10.1038/s41467-025-57982-5} {\bibfield  {journal} {\bibinfo  {journal} {Nat. Commun.}\ }\textbf {\bibinfo {volume} {16}},\ \bibinfo {pages} {3290} (\bibinfo {year} {2025})}\BibitemShut {NoStop}%
\bibitem [{\citenamefont {Persat}\ \emph {et~al.}(2015)\citenamefont {Persat}, \citenamefont {Nadell}, \citenamefont {Kim}, \citenamefont {Ingremeau}, \citenamefont {Siryaporn}, \citenamefont {Drescher}, \citenamefont {Wingreen}, \citenamefont {Bassler}, \citenamefont {Gitai},\ and\ \citenamefont {Stone}}]{persat2015mechanical}%
  \BibitemOpen
  \bibfield  {author} {\bibinfo {author} {\bibfnamefont {A.}~\bibnamefont {Persat}}, \bibinfo {author} {\bibfnamefont {C.~D.}\ \bibnamefont {Nadell}}, \bibinfo {author} {\bibfnamefont {M.~K.}\ \bibnamefont {Kim}}, \bibinfo {author} {\bibfnamefont {F.}~\bibnamefont {Ingremeau}}, \bibinfo {author} {\bibfnamefont {A.}~\bibnamefont {Siryaporn}}, \bibinfo {author} {\bibfnamefont {K.}~\bibnamefont {Drescher}}, \bibinfo {author} {\bibfnamefont {N.~S.}\ \bibnamefont {Wingreen}}, \bibinfo {author} {\bibfnamefont {B.~L.}\ \bibnamefont {Bassler}}, \bibinfo {author} {\bibfnamefont {Z.}~\bibnamefont {Gitai}},\ and\ \bibinfo {author} {\bibfnamefont {H.~A.}\ \bibnamefont {Stone}},\ }\href {https://doi.org/10.1016/j.cell.2015.05.005} {\bibfield  {journal} {\bibinfo  {journal} {Cell}\ }\textbf {\bibinfo {volume} {161}},\ \bibinfo {pages} {988} (\bibinfo {year} {2015})}\BibitemShut {NoStop}%
\bibitem [{\citenamefont {Dufr{\^e}ne}\ and\ \citenamefont {Persat}(2020)}]{dufrene2020mechanomicrobiology}%
  \BibitemOpen
  \bibfield  {author} {\bibinfo {author} {\bibfnamefont {Y.~F.}\ \bibnamefont {Dufr{\^e}ne}}\ and\ \bibinfo {author} {\bibfnamefont {A.}~\bibnamefont {Persat}},\ }\href {https://doi.org/10.1038/s41579-019-0314-2} {\bibfield  {journal} {\bibinfo  {journal} {Nat. Rev. Microbiol.}\ }\textbf {\bibinfo {volume} {18}},\ \bibinfo {pages} {227} (\bibinfo {year} {2020})}\BibitemShut {NoStop}%
\bibitem [{\citenamefont {Kimkes}\ and\ \citenamefont {Heinemann}(2020)}]{kimkes2020bacteria}%
  \BibitemOpen
  \bibfield  {author} {\bibinfo {author} {\bibfnamefont {T.~E.}\ \bibnamefont {Kimkes}}\ and\ \bibinfo {author} {\bibfnamefont {M.}~\bibnamefont {Heinemann}},\ }\href {https://doi.org/10.1093/femsre/fuz029} {\bibfield  {journal} {\bibinfo  {journal} {FEMS Microbiol. Rev.}\ }\textbf {\bibinfo {volume} {44}},\ \bibinfo {pages} {106} (\bibinfo {year} {2020})}\BibitemShut {NoStop}%
\bibitem [{\citenamefont {Siryaporn}\ \emph {et~al.}(2014)\citenamefont {Siryaporn}, \citenamefont {Kuchma}, \citenamefont {O'Toole},\ and\ \citenamefont {Gitai}}]{siryaporn2014surface}%
  \BibitemOpen
  \bibfield  {author} {\bibinfo {author} {\bibfnamefont {A.}~\bibnamefont {Siryaporn}}, \bibinfo {author} {\bibfnamefont {S.~L.}\ \bibnamefont {Kuchma}}, \bibinfo {author} {\bibfnamefont {G.~A.}\ \bibnamefont {O'Toole}},\ and\ \bibinfo {author} {\bibfnamefont {Z.}~\bibnamefont {Gitai}},\ }\href {https://doi.org/10.1073/pnas.1415712111} {\bibfield  {journal} {\bibinfo  {journal} {Proc. Natl. Acad. Sci. USA}\ }\textbf {\bibinfo {volume} {111}},\ \bibinfo {pages} {16860} (\bibinfo {year} {2014})}\BibitemShut {NoStop}%
\bibitem [{\citenamefont {Zhou}\ \emph {et~al.}(2025)\citenamefont {Zhou}, \citenamefont {Zhou}, \citenamefont {Zhang}, \citenamefont {Zhao}, \citenamefont {Ye}, \citenamefont {Xu},\ and\ \citenamefont {Li}}]{zhou2025confined}%
  \BibitemOpen
  \bibfield  {author} {\bibinfo {author} {\bibfnamefont {Y.}~\bibnamefont {Zhou}}, \bibinfo {author} {\bibfnamefont {X.}~\bibnamefont {Zhou}}, \bibinfo {author} {\bibfnamefont {J.}~\bibnamefont {Zhang}}, \bibinfo {author} {\bibfnamefont {Y.}~\bibnamefont {Zhao}}, \bibinfo {author} {\bibfnamefont {Z.}~\bibnamefont {Ye}}, \bibinfo {author} {\bibfnamefont {F.}~\bibnamefont {Xu}},\ and\ \bibinfo {author} {\bibfnamefont {F.}~\bibnamefont {Li}},\ }\href {https://doi.org/10.1021/acs.analchem.4c05503} {\bibfield  {journal} {\bibinfo  {journal} {Anal. Chem.}\ }\textbf {\bibinfo {volume} {97}},\ \bibinfo {pages} {5517} (\bibinfo {year} {2025})}\BibitemShut {NoStop}%
\bibitem [{\citenamefont {Pelech}\ \emph {et~al.}(2025)\citenamefont {Pelech}, \citenamefont {Navarro}, \citenamefont {Vettiger}, \citenamefont {Chao},\ and\ \citenamefont {Allolio}}]{pelech2025stress}%
  \BibitemOpen
  \bibfield  {author} {\bibinfo {author} {\bibfnamefont {P.}~\bibnamefont {Pelech}}, \bibinfo {author} {\bibfnamefont {P.~P.}\ \bibnamefont {Navarro}}, \bibinfo {author} {\bibfnamefont {A.}~\bibnamefont {Vettiger}}, \bibinfo {author} {\bibfnamefont {L.~H.}\ \bibnamefont {Chao}},\ and\ \bibinfo {author} {\bibfnamefont {C.}~\bibnamefont {Allolio}},\ }\href {https://doi.org/10.1073/pnas.2424441122} {\bibfield  {journal} {\bibinfo  {journal} {Proc. Natl. Acad. Sci. USA}\ }\textbf {\bibinfo {volume} {122}},\ \bibinfo {pages} {e2424441122} (\bibinfo {year} {2025})}\BibitemShut {NoStop}%
\bibitem [{\citenamefont {Tuson}\ \emph {et~al.}(2012)\citenamefont {Tuson}, \citenamefont {Auer}, \citenamefont {Renner}, \citenamefont {Hasebe}, \citenamefont {Tropini}, \citenamefont {Salick}, \citenamefont {Crone}, \citenamefont {Gopinathan}, \citenamefont {Huang},\ and\ \citenamefont {Weibel}}]{tuson2012measuring}%
  \BibitemOpen
  \bibfield  {author} {\bibinfo {author} {\bibfnamefont {H.~H.}\ \bibnamefont {Tuson}}, \bibinfo {author} {\bibfnamefont {G.~K.}\ \bibnamefont {Auer}}, \bibinfo {author} {\bibfnamefont {L.~D.}\ \bibnamefont {Renner}}, \bibinfo {author} {\bibfnamefont {M.}~\bibnamefont {Hasebe}}, \bibinfo {author} {\bibfnamefont {C.}~\bibnamefont {Tropini}}, \bibinfo {author} {\bibfnamefont {M.}~\bibnamefont {Salick}}, \bibinfo {author} {\bibfnamefont {W.~C.}\ \bibnamefont {Crone}}, \bibinfo {author} {\bibfnamefont {A.}~\bibnamefont {Gopinathan}}, \bibinfo {author} {\bibfnamefont {K.~C.}\ \bibnamefont {Huang}},\ and\ \bibinfo {author} {\bibfnamefont {D.~B.}\ \bibnamefont {Weibel}},\ }\href {https://doi.org/10.1111/j.1365-2958.2012.08063.x} {\bibfield  {journal} {\bibinfo  {journal} {Mol. Microbiol.}\ }\textbf {\bibinfo {volume} {84}},\ \bibinfo {pages} {874} (\bibinfo {year} {2012})}\BibitemShut {NoStop}%
\bibitem [{\citenamefont {Wittmann}\ \emph {et~al.}(2023)\citenamefont {Wittmann}, \citenamefont {Nguyen}, \citenamefont {L{\"o}wen}, \citenamefont {Schwarzendahl},\ and\ \citenamefont {Sengupta}}]{wittmann2023collective}%
  \BibitemOpen
  \bibfield  {author} {\bibinfo {author} {\bibfnamefont {R.}~\bibnamefont {Wittmann}}, \bibinfo {author} {\bibfnamefont {G.~P.}\ \bibnamefont {Nguyen}}, \bibinfo {author} {\bibfnamefont {H.}~\bibnamefont {L{\"o}wen}}, \bibinfo {author} {\bibfnamefont {F.~J.}\ \bibnamefont {Schwarzendahl}},\ and\ \bibinfo {author} {\bibfnamefont {A.}~\bibnamefont {Sengupta}},\ }\href {https://doi.org/10.1038/s42005-023-01449-w} {\bibfield  {journal} {\bibinfo  {journal} {Commun. Phys.}\ }\textbf {\bibinfo {volume} {6}},\ \bibinfo {pages} {331} (\bibinfo {year} {2023})}\BibitemShut {NoStop}%
\bibitem [{\citenamefont {Chu}\ \emph {et~al.}(2018)\citenamefont {Chu}, \citenamefont {Kilic}, \citenamefont {Cho}, \citenamefont {Groisman},\ and\ \citenamefont {Levchenko}}]{chu2018self}%
  \BibitemOpen
  \bibfield  {author} {\bibinfo {author} {\bibfnamefont {E.~K.}\ \bibnamefont {Chu}}, \bibinfo {author} {\bibfnamefont {O.}~\bibnamefont {Kilic}}, \bibinfo {author} {\bibfnamefont {H.}~\bibnamefont {Cho}}, \bibinfo {author} {\bibfnamefont {A.}~\bibnamefont {Groisman}},\ and\ \bibinfo {author} {\bibfnamefont {A.}~\bibnamefont {Levchenko}},\ }\href {https://doi.org/10.1038/s41467-018-06552-z} {\bibfield  {journal} {\bibinfo  {journal} {Nat. Commun.}\ }\textbf {\bibinfo {volume} {9}},\ \bibinfo {pages} {4087} (\bibinfo {year} {2018})}\BibitemShut {NoStop}%
\bibitem [{\citenamefont {Kumar}\ and\ \citenamefont {Libchaber}(2013)}]{kumar2013pressure}%
  \BibitemOpen
  \bibfield  {author} {\bibinfo {author} {\bibfnamefont {P.}~\bibnamefont {Kumar}}\ and\ \bibinfo {author} {\bibfnamefont {A.}~\bibnamefont {Libchaber}},\ }\href {https://doi.org/10.1016/j.bpj.2013.06.029} {\bibfield  {journal} {\bibinfo  {journal} {Biophys. J.}\ }\textbf {\bibinfo {volume} {105}},\ \bibinfo {pages} {783} (\bibinfo {year} {2013})}\BibitemShut {NoStop}%
\bibitem [{\citenamefont {Nepal}\ and\ \citenamefont {Kumar}(2018)}]{nepal2018dynamics}%
  \BibitemOpen
  \bibfield  {author} {\bibinfo {author} {\bibfnamefont {S.}~\bibnamefont {Nepal}}\ and\ \bibinfo {author} {\bibfnamefont {P.}~\bibnamefont {Kumar}},\ }\href {https://doi.org/10.1103/PhysRevE.97.052411} {\bibfield  {journal} {\bibinfo  {journal} {Phys. Rev. E}\ }\textbf {\bibinfo {volume} {97}},\ \bibinfo {pages} {052411} (\bibinfo {year} {2018})}\BibitemShut {NoStop}%
\bibitem [{\citenamefont {Le~Blanc}\ \emph {et~al.}(2026)\citenamefont {Le~Blanc}, \citenamefont {Alric}, \citenamefont {Rollin}, \citenamefont {X{\'e}nard}, \citenamefont {Avdovic}, \citenamefont {Goussard}, \citenamefont {Mazenq}, \citenamefont {Piel}, \citenamefont {Tinevez}, \citenamefont {Delarue}, \citenamefont {Dum{\'e}nil},\ and\ \citenamefont {Bonazzi}}]{leblanc2024bacterial}%
  \BibitemOpen
  \bibfield  {author} {\bibinfo {author} {\bibfnamefont {L.}~\bibnamefont {Le~Blanc}}, \bibinfo {author} {\bibfnamefont {B.}~\bibnamefont {Alric}}, \bibinfo {author} {\bibfnamefont {R.}~\bibnamefont {Rollin}}, \bibinfo {author} {\bibfnamefont {L.}~\bibnamefont {X{\'e}nard}}, \bibinfo {author} {\bibfnamefont {M.}~\bibnamefont {Avdovic}}, \bibinfo {author} {\bibfnamefont {S.}~\bibnamefont {Goussard}}, \bibinfo {author} {\bibfnamefont {L.}~\bibnamefont {Mazenq}}, \bibinfo {author} {\bibfnamefont {M.}~\bibnamefont {Piel}}, \bibinfo {author} {\bibfnamefont {J.-Y.}\ \bibnamefont {Tinevez}}, \bibinfo {author} {\bibfnamefont {M.}~\bibnamefont {Delarue}}, \bibinfo {author} {\bibfnamefont {G.}~\bibnamefont {Dum{\'e}nil}},\ and\ \bibinfo {author} {\bibfnamefont {D.}~\bibnamefont {Bonazzi}},\ }\bibfield  {journal} {\bibinfo  {journal} {Nat. Commun.}\ }\href {https://doi.org/10.1038/s41467-026-76492-6} {10.1038/s41467-026-76492-6} (\bibinfo {year} {2026})\BibitemShut {NoStop}%
\bibitem [{\citenamefont {Aguilar-Luviano}\ \emph {et~al.}(2026)\citenamefont {Aguilar-Luviano}, \citenamefont {Santos-Escobar}, \citenamefont {Orozco-Barrera},\ and\ \citenamefont {Pe{\~n}a-Miller}}]{luviano2026conditional}%
  \BibitemOpen
  \bibfield  {author} {\bibinfo {author} {\bibfnamefont {O.~B.}\ \bibnamefont {Aguilar-Luviano}}, \bibinfo {author} {\bibfnamefont {F.}~\bibnamefont {Santos-Escobar}}, \bibinfo {author} {\bibfnamefont {S.}~\bibnamefont {Orozco-Barrera}},\ and\ \bibinfo {author} {\bibfnamefont {R.}~\bibnamefont {Pe{\~n}a-Miller}},\ }\href {https://doi.org/10.1098/rsif.2025.0959} {\bibfield  {journal} {\bibinfo  {journal} {J. R. Soc. Interface}\ }\textbf {\bibinfo {volume} {23}},\ \bibinfo {pages} {20250959} (\bibinfo {year} {2026})}\BibitemShut {NoStop}%
\bibitem [{\citenamefont {Cayron}\ \emph {et~al.}(2023)\citenamefont {Cayron}, \citenamefont {Dedieu-Berne},\ and\ \citenamefont {Lesterlin}}]{cayron2023bacterial}%
  \BibitemOpen
  \bibfield  {author} {\bibinfo {author} {\bibfnamefont {J.}~\bibnamefont {Cayron}}, \bibinfo {author} {\bibfnamefont {A.}~\bibnamefont {Dedieu-Berne}},\ and\ \bibinfo {author} {\bibfnamefont {C.}~\bibnamefont {Lesterlin}},\ }\href {https://doi.org/10.1111/mmi.15016} {\bibfield  {journal} {\bibinfo  {journal} {Mol. Microbiol.}\ }\textbf {\bibinfo {volume} {119}},\ \bibinfo {pages} {237} (\bibinfo {year} {2023})}\BibitemShut {NoStop}%
\bibitem [{\citenamefont {M{\"a}nnik}\ \emph {et~al.}(2009)\citenamefont {M{\"a}nnik}, \citenamefont {Driessen}, \citenamefont {Galajda}, \citenamefont {Keymer},\ and\ \citenamefont {Dekker}}]{mannik2009bacterial}%
  \BibitemOpen
  \bibfield  {author} {\bibinfo {author} {\bibfnamefont {J.}~\bibnamefont {M{\"a}nnik}}, \bibinfo {author} {\bibfnamefont {R.}~\bibnamefont {Driessen}}, \bibinfo {author} {\bibfnamefont {P.}~\bibnamefont {Galajda}}, \bibinfo {author} {\bibfnamefont {J.~E.}\ \bibnamefont {Keymer}},\ and\ \bibinfo {author} {\bibfnamefont {C.}~\bibnamefont {Dekker}},\ }\href {https://doi.org/10.1073/pnas.0907542106} {\bibfield  {journal} {\bibinfo  {journal} {Proc. Natl. Acad. Sci. USA}\ }\textbf {\bibinfo {volume} {106}},\ \bibinfo {pages} {14861} (\bibinfo {year} {2009})}\BibitemShut {NoStop}%
\bibitem [{\citenamefont {Si}\ \emph {et~al.}(2015)\citenamefont {Si}, \citenamefont {Li}, \citenamefont {Margolin},\ and\ \citenamefont {Sun}}]{si2015bacterial}%
  \BibitemOpen
  \bibfield  {author} {\bibinfo {author} {\bibfnamefont {F.}~\bibnamefont {Si}}, \bibinfo {author} {\bibfnamefont {B.}~\bibnamefont {Li}}, \bibinfo {author} {\bibfnamefont {W.}~\bibnamefont {Margolin}},\ and\ \bibinfo {author} {\bibfnamefont {S.~X.}\ \bibnamefont {Sun}},\ }\href {https://doi.org/10.1038/srep11367} {\bibfield  {journal} {\bibinfo  {journal} {Sci. Rep.}\ }\textbf {\bibinfo {volume} {5}},\ \bibinfo {pages} {11367} (\bibinfo {year} {2015})}\BibitemShut {NoStop}%
\bibitem [{\citenamefont {Ma{\~n}as}\ and\ \citenamefont {Mackey}(2004)}]{manas2004morphological}%
  \BibitemOpen
  \bibfield  {author} {\bibinfo {author} {\bibfnamefont {P.}~\bibnamefont {Ma{\~n}as}}\ and\ \bibinfo {author} {\bibfnamefont {B.~M.}\ \bibnamefont {Mackey}},\ }\href {https://doi.org/10.1128/AEM.70.3.1545-1554.2004} {\bibfield  {journal} {\bibinfo  {journal} {Appl. Environ. Microbiol.}\ }\textbf {\bibinfo {volume} {70}},\ \bibinfo {pages} {1545} (\bibinfo {year} {2004})}\BibitemShut {NoStop}%
\bibitem [{\citenamefont {Allen}\ and\ \citenamefont {Waclaw}(2019)}]{allen2019bacterial}%
  \BibitemOpen
  \bibfield  {author} {\bibinfo {author} {\bibfnamefont {R.~J.}\ \bibnamefont {Allen}}\ and\ \bibinfo {author} {\bibfnamefont {B.}~\bibnamefont {Waclaw}},\ }\href {https://doi.org/10.1088/1361-6633/aae546} {\bibfield  {journal} {\bibinfo  {journal} {Rep. Prog. Phys.}\ }\textbf {\bibinfo {volume} {82}},\ \bibinfo {pages} {016601} (\bibinfo {year} {2019})}\BibitemShut {NoStop}%
\bibitem [{\citenamefont {Ara{\'u}jo}\ \emph {et~al.}(2023)\citenamefont {Ara{\'u}jo}, \citenamefont {Janssen}, \citenamefont {Barois}, \citenamefont {Boffetta}, \citenamefont {Cohen}, \citenamefont {Corbetta}, \citenamefont {Dauchot}, \citenamefont {Dijkstra}, \citenamefont {Durham}, \citenamefont {Dussutour} \emph {et~al.}}]{araujo2023steering}%
  \BibitemOpen
  \bibfield  {author} {\bibinfo {author} {\bibfnamefont {N.~A.}\ \bibnamefont {Ara{\'u}jo}}, \bibinfo {author} {\bibfnamefont {L.~M.}\ \bibnamefont {Janssen}}, \bibinfo {author} {\bibfnamefont {T.}~\bibnamefont {Barois}}, \bibinfo {author} {\bibfnamefont {G.}~\bibnamefont {Boffetta}}, \bibinfo {author} {\bibfnamefont {I.}~\bibnamefont {Cohen}}, \bibinfo {author} {\bibfnamefont {A.}~\bibnamefont {Corbetta}}, \bibinfo {author} {\bibfnamefont {O.}~\bibnamefont {Dauchot}}, \bibinfo {author} {\bibfnamefont {M.}~\bibnamefont {Dijkstra}}, \bibinfo {author} {\bibfnamefont {W.~M.}\ \bibnamefont {Durham}}, \bibinfo {author} {\bibfnamefont {A.}~\bibnamefont {Dussutour}}, \emph {et~al.},\ }\href {https://doi.org/10.1039/D2SM01562E} {\bibfield  {journal} {\bibinfo  {journal} {Soft Matter}\ }\textbf {\bibinfo {volume} {19}},\ \bibinfo {pages} {1695} (\bibinfo {year} {2023})}\BibitemShut {NoStop}%
\bibitem [{\citenamefont {Meacock}\ \emph {et~al.}(2021)\citenamefont {Meacock}, \citenamefont {Doostmohammadi}, \citenamefont {Foster}, \citenamefont {Yeomans},\ and\ \citenamefont {Durham}}]{meacock2021bacteria}%
  \BibitemOpen
  \bibfield  {author} {\bibinfo {author} {\bibfnamefont {O.~J.}\ \bibnamefont {Meacock}}, \bibinfo {author} {\bibfnamefont {A.}~\bibnamefont {Doostmohammadi}}, \bibinfo {author} {\bibfnamefont {K.~R.}\ \bibnamefont {Foster}}, \bibinfo {author} {\bibfnamefont {J.~M.}\ \bibnamefont {Yeomans}},\ and\ \bibinfo {author} {\bibfnamefont {W.~M.}\ \bibnamefont {Durham}},\ }\href {https://doi.org/10.1038/s41567-020-01070-6} {\bibfield  {journal} {\bibinfo  {journal} {Nat. Phys.}\ }\textbf {\bibinfo {volume} {17}},\ \bibinfo {pages} {205} (\bibinfo {year} {2021})}\BibitemShut {NoStop}%
\bibitem [{\citenamefont {Sengupta}(2020)}]{sengupta2020microbial}%
  \BibitemOpen
  \bibfield  {author} {\bibinfo {author} {\bibfnamefont {A.}~\bibnamefont {Sengupta}},\ }\href {https://doi.org/10.3389/fphy.2020.00184} {\bibfield  {journal} {\bibinfo  {journal} {Front. Phys.}\ }\textbf {\bibinfo {volume} {8}},\ \bibinfo {pages} {184} (\bibinfo {year} {2020})}\BibitemShut {NoStop}%
\bibitem [{\citenamefont {Dhar}\ \emph {et~al.}(2022)\citenamefont {Dhar}, \citenamefont {Thai}, \citenamefont {Ghoshal}, \citenamefont {Giomi},\ and\ \citenamefont {Sengupta}}]{dhar2022self}%
  \BibitemOpen
  \bibfield  {author} {\bibinfo {author} {\bibfnamefont {J.}~\bibnamefont {Dhar}}, \bibinfo {author} {\bibfnamefont {A.~L.}\ \bibnamefont {Thai}}, \bibinfo {author} {\bibfnamefont {A.}~\bibnamefont {Ghoshal}}, \bibinfo {author} {\bibfnamefont {L.}~\bibnamefont {Giomi}},\ and\ \bibinfo {author} {\bibfnamefont {A.}~\bibnamefont {Sengupta}},\ }\href {https://doi.org/10.1038/s41567-022-01641-9} {\bibfield  {journal} {\bibinfo  {journal} {Nat. Phys.}\ }\textbf {\bibinfo {volume} {18}},\ \bibinfo {pages} {945} (\bibinfo {year} {2022})}\BibitemShut {NoStop}%
\bibitem [{\citenamefont {Rani}\ \emph {et~al.}()\citenamefont {Rani}, \citenamefont {Nguyen}, \citenamefont {Wittmann}, \citenamefont {L{\"o}wen},\ and\ \citenamefont {Sengupta}}]{rani2025nascent}%
  \BibitemOpen
  \bibfield  {author} {\bibinfo {author} {\bibfnamefont {G.}~\bibnamefont {Rani}}, \bibinfo {author} {\bibfnamefont {G.~H.~P.}\ \bibnamefont {Nguyen}}, \bibinfo {author} {\bibfnamefont {R.}~\bibnamefont {Wittmann}}, \bibinfo {author} {\bibfnamefont {H.}~\bibnamefont {L{\"o}wen}},\ and\ \bibinfo {author} {\bibfnamefont {A.}~\bibnamefont {Sengupta}},\ }\href@noop {} {\bibinfo {title} {Nascent biofilms on soft surfaces}},\ \Eprint {https://arxiv.org/abs/2512.21870} {arXiv:2512.21870 [physics.bio-ph]} \BibitemShut {NoStop}%
\bibitem [{\citenamefont {Sreepadmanabh}\ \emph {et~al.}(2024)\citenamefont {Sreepadmanabh}, \citenamefont {Ganesh}, \citenamefont {Sanjenbam}, \citenamefont {Kurzthaler}, \citenamefont {Agashe},\ and\ \citenamefont {Bhattacharjee}}]{sreepadmanabh2024cell}%
  \BibitemOpen
  \bibfield  {author} {\bibinfo {author} {\bibfnamefont {M.}~\bibnamefont {Sreepadmanabh}}, \bibinfo {author} {\bibfnamefont {M.}~\bibnamefont {Ganesh}}, \bibinfo {author} {\bibfnamefont {P.}~\bibnamefont {Sanjenbam}}, \bibinfo {author} {\bibfnamefont {C.}~\bibnamefont {Kurzthaler}}, \bibinfo {author} {\bibfnamefont {D.}~\bibnamefont {Agashe}},\ and\ \bibinfo {author} {\bibfnamefont {T.}~\bibnamefont {Bhattacharjee}},\ }\href {https://doi.org/10.1038/s41467-024-53989-6} {\bibfield  {journal} {\bibinfo  {journal} {Nat. Commun.}\ }\textbf {\bibinfo {volume} {15}},\ \bibinfo {pages} {9561} (\bibinfo {year} {2024})}\BibitemShut {NoStop}%
\bibitem [{\citenamefont {Hancock}\ and\ \citenamefont {Datta}(2024)}]{hancock2024interplay}%
  \BibitemOpen
  \bibfield  {author} {\bibinfo {author} {\bibfnamefont {A.~M.}\ \bibnamefont {Hancock}}\ and\ \bibinfo {author} {\bibfnamefont {S.~S.}\ \bibnamefont {Datta}},\ }\href {https://doi.org/10.1016/j.bpj.2024.03.008} {\bibfield  {journal} {\bibinfo  {journal} {Biophys. J.}\ }\textbf {\bibinfo {volume} {123}},\ \bibinfo {pages} {957} (\bibinfo {year} {2024})}\BibitemShut {NoStop}%
\bibitem [{\citenamefont {Zimmer}\ \emph {et~al.}(2026)\citenamefont {Zimmer}, \citenamefont {Bittihn},\ and\ \citenamefont {Pollack}}]{zimmer2026dimensionality}%
  \BibitemOpen
  \bibfield  {author} {\bibinfo {author} {\bibfnamefont {P.}~\bibnamefont {Zimmer}}, \bibinfo {author} {\bibfnamefont {P.}~\bibnamefont {Bittihn}},\ and\ \bibinfo {author} {\bibfnamefont {Y.~G.}\ \bibnamefont {Pollack}},\ }\href {https://doi.org/10.1088/1367-2630/ae7ea6} {\bibfield  {journal} {\bibinfo  {journal} {New J. Phys.}\ }\textbf {\bibinfo {volume} {28}},\ \bibinfo {pages} {084402} (\bibinfo {year} {2026})}\BibitemShut {NoStop}%
\bibitem [{\citenamefont {Jun}\ \emph {et~al.}(2018)\citenamefont {Jun}, \citenamefont {Si}, \citenamefont {Pugatch},\ and\ \citenamefont {Scott}}]{jun2018fundamental}%
  \BibitemOpen
  \bibfield  {author} {\bibinfo {author} {\bibfnamefont {S.}~\bibnamefont {Jun}}, \bibinfo {author} {\bibfnamefont {F.}~\bibnamefont {Si}}, \bibinfo {author} {\bibfnamefont {R.}~\bibnamefont {Pugatch}},\ and\ \bibinfo {author} {\bibfnamefont {M.}~\bibnamefont {Scott}},\ }\href {https://doi.org/10.1088/1361-6633/aaa628} {\bibfield  {journal} {\bibinfo  {journal} {Rep. Prog. Phys.}\ }\textbf {\bibinfo {volume} {81}},\ \bibinfo {pages} {056601} (\bibinfo {year} {2018})}\BibitemShut {NoStop}%
\bibitem [{\citenamefont {Wang}\ \emph {et~al.}(2010)\citenamefont {Wang}, \citenamefont {Robert}, \citenamefont {Pelletier}, \citenamefont {Dang}, \citenamefont {Taddei}, \citenamefont {Wright},\ and\ \citenamefont {Jun}}]{wang2010robust}%
  \BibitemOpen
  \bibfield  {author} {\bibinfo {author} {\bibfnamefont {P.}~\bibnamefont {Wang}}, \bibinfo {author} {\bibfnamefont {L.}~\bibnamefont {Robert}}, \bibinfo {author} {\bibfnamefont {J.}~\bibnamefont {Pelletier}}, \bibinfo {author} {\bibfnamefont {W.~L.}\ \bibnamefont {Dang}}, \bibinfo {author} {\bibfnamefont {F.}~\bibnamefont {Taddei}}, \bibinfo {author} {\bibfnamefont {A.}~\bibnamefont {Wright}},\ and\ \bibinfo {author} {\bibfnamefont {S.}~\bibnamefont {Jun}},\ }\href {https://doi.org/10.1016/j.cub.2010.04.045} {\bibfield  {journal} {\bibinfo  {journal} {Curr. Biol.}\ }\textbf {\bibinfo {volume} {20}},\ \bibinfo {pages} {1099} (\bibinfo {year} {2010})}\BibitemShut {NoStop}%
\bibitem [{\citenamefont {Campos}\ \emph {et~al.}(2014)\citenamefont {Campos}, \citenamefont {Surovtsev}, \citenamefont {Kato}, \citenamefont {Paintdakhi}, \citenamefont {Beltran}, \citenamefont {Ebmeier},\ and\ \citenamefont {Jacobs-Wagner}}]{campos2014constant}%
  \BibitemOpen
  \bibfield  {author} {\bibinfo {author} {\bibfnamefont {M.}~\bibnamefont {Campos}}, \bibinfo {author} {\bibfnamefont {I.~V.}\ \bibnamefont {Surovtsev}}, \bibinfo {author} {\bibfnamefont {S.}~\bibnamefont {Kato}}, \bibinfo {author} {\bibfnamefont {A.}~\bibnamefont {Paintdakhi}}, \bibinfo {author} {\bibfnamefont {B.}~\bibnamefont {Beltran}}, \bibinfo {author} {\bibfnamefont {S.~E.}\ \bibnamefont {Ebmeier}},\ and\ \bibinfo {author} {\bibfnamefont {C.}~\bibnamefont {Jacobs-Wagner}},\ }\href {https://doi.org/10.1016/j.cell.2014.11.022} {\bibfield  {journal} {\bibinfo  {journal} {Cell}\ }\textbf {\bibinfo {volume} {159}},\ \bibinfo {pages} {1433} (\bibinfo {year} {2014})}\BibitemShut {NoStop}%
\bibitem [{\citenamefont {Taheri-Araghi}\ \emph {et~al.}(2015)\citenamefont {Taheri-Araghi}, \citenamefont {Bradde}, \citenamefont {Sauls}, \citenamefont {Hill}, \citenamefont {Levin}, \citenamefont {Paulsson}, \citenamefont {Vergassola},\ and\ \citenamefont {Jun}}]{taheri2015cell}%
  \BibitemOpen
  \bibfield  {author} {\bibinfo {author} {\bibfnamefont {S.}~\bibnamefont {Taheri-Araghi}}, \bibinfo {author} {\bibfnamefont {S.}~\bibnamefont {Bradde}}, \bibinfo {author} {\bibfnamefont {J.~T.}\ \bibnamefont {Sauls}}, \bibinfo {author} {\bibfnamefont {N.~S.}\ \bibnamefont {Hill}}, \bibinfo {author} {\bibfnamefont {P.~A.}\ \bibnamefont {Levin}}, \bibinfo {author} {\bibfnamefont {J.}~\bibnamefont {Paulsson}}, \bibinfo {author} {\bibfnamefont {M.}~\bibnamefont {Vergassola}},\ and\ \bibinfo {author} {\bibfnamefont {S.}~\bibnamefont {Jun}},\ }\href {https://doi.org/10.1016/j.cub.2014.12.009} {\bibfield  {journal} {\bibinfo  {journal} {Curr. Biol.}\ }\textbf {\bibinfo {volume} {25}},\ \bibinfo {pages} {385} (\bibinfo {year} {2015})}\BibitemShut {NoStop}%
\bibitem [{\citenamefont {Kennard}\ \emph {et~al.}(2016)\citenamefont {Kennard}, \citenamefont {Osella}, \citenamefont {Javer}, \citenamefont {Grilli}, \citenamefont {Nghe}, \citenamefont {Tans}, \citenamefont {Cicuta},\ and\ \citenamefont {Cosentino~Lagomarsino}}]{kennard2016individuality}%
  \BibitemOpen
  \bibfield  {author} {\bibinfo {author} {\bibfnamefont {A.~S.}\ \bibnamefont {Kennard}}, \bibinfo {author} {\bibfnamefont {M.}~\bibnamefont {Osella}}, \bibinfo {author} {\bibfnamefont {A.}~\bibnamefont {Javer}}, \bibinfo {author} {\bibfnamefont {J.}~\bibnamefont {Grilli}}, \bibinfo {author} {\bibfnamefont {P.}~\bibnamefont {Nghe}}, \bibinfo {author} {\bibfnamefont {S.~J.}\ \bibnamefont {Tans}}, \bibinfo {author} {\bibfnamefont {P.}~\bibnamefont {Cicuta}},\ and\ \bibinfo {author} {\bibfnamefont {M.}~\bibnamefont {Cosentino~Lagomarsino}},\ }\href {https://doi.org/10.1103/PhysRevE.93.012408} {\bibfield  {journal} {\bibinfo  {journal} {Phys. Rev. E}\ }\textbf {\bibinfo {volume} {93}},\ \bibinfo {pages} {012408} (\bibinfo {year} {2016})}\BibitemShut {NoStop}%
\bibitem [{\citenamefont {Si}\ \emph {et~al.}(2019)\citenamefont {Si}, \citenamefont {Le~Treut}, \citenamefont {Sauls}, \citenamefont {Vadia}, \citenamefont {Levin},\ and\ \citenamefont {Jun}}]{si2019mechanistic}%
  \BibitemOpen
  \bibfield  {author} {\bibinfo {author} {\bibfnamefont {F.}~\bibnamefont {Si}}, \bibinfo {author} {\bibfnamefont {G.}~\bibnamefont {Le~Treut}}, \bibinfo {author} {\bibfnamefont {J.~T.}\ \bibnamefont {Sauls}}, \bibinfo {author} {\bibfnamefont {S.}~\bibnamefont {Vadia}}, \bibinfo {author} {\bibfnamefont {P.~A.}\ \bibnamefont {Levin}},\ and\ \bibinfo {author} {\bibfnamefont {S.}~\bibnamefont {Jun}},\ }\href {https://doi.org/10.1016/j.cub.2019.04.062} {\bibfield  {journal} {\bibinfo  {journal} {Curr. Biol.}\ }\textbf {\bibinfo {volume} {29}},\ \bibinfo {pages} {1760} (\bibinfo {year} {2019})}\BibitemShut {NoStop}%
\bibitem [{\citenamefont {Miguel}\ \emph {et~al.}(2025)\citenamefont {Miguel}, \citenamefont {Zietek}, \citenamefont {Shi}, \citenamefont {Sueki}, \citenamefont {Corona}, \citenamefont {Maier}, \citenamefont {Verheul}, \citenamefont {den Blaauwen}, \citenamefont {Van~Valen}, \citenamefont {Typas} \emph {et~al.}}]{miguel2025modulation}%
  \BibitemOpen
  \bibfield  {author} {\bibinfo {author} {\bibfnamefont {A.}~\bibnamefont {Miguel}}, \bibinfo {author} {\bibfnamefont {M.}~\bibnamefont {Zietek}}, \bibinfo {author} {\bibfnamefont {H.}~\bibnamefont {Shi}}, \bibinfo {author} {\bibfnamefont {A.}~\bibnamefont {Sueki}}, \bibinfo {author} {\bibfnamefont {F.}~\bibnamefont {Corona}}, \bibinfo {author} {\bibfnamefont {L.}~\bibnamefont {Maier}}, \bibinfo {author} {\bibfnamefont {J.}~\bibnamefont {Verheul}}, \bibinfo {author} {\bibfnamefont {T.}~\bibnamefont {den Blaauwen}}, \bibinfo {author} {\bibfnamefont {D.}~\bibnamefont {Van~Valen}}, \bibinfo {author} {\bibfnamefont {A.}~\bibnamefont {Typas}}, \emph {et~al.},\ }\href {https://doi.org/10.1128/mbio.02281-25} {\bibfield  {journal} {\bibinfo  {journal} {mBio}\ }\textbf {\bibinfo {volume} {16}},\ \bibinfo {pages} {e02281} (\bibinfo {year} {2025})}\BibitemShut {NoStop}%
\bibitem [{\citenamefont {You}\ \emph {et~al.}(2018)\citenamefont {You}, \citenamefont {Pearce}, \citenamefont {Sengupta},\ and\ \citenamefont {Giomi}}]{you2018geometry}%
  \BibitemOpen
  \bibfield  {author} {\bibinfo {author} {\bibfnamefont {Z.}~\bibnamefont {You}}, \bibinfo {author} {\bibfnamefont {D.~J.}\ \bibnamefont {Pearce}}, \bibinfo {author} {\bibfnamefont {A.}~\bibnamefont {Sengupta}},\ and\ \bibinfo {author} {\bibfnamefont {L.}~\bibnamefont {Giomi}},\ }\href {https://doi.org/10.1103/PhysRevX.8.031065} {\bibfield  {journal} {\bibinfo  {journal} {Phys. Rev. X}\ }\textbf {\bibinfo {volume} {8}},\ \bibinfo {pages} {031065} (\bibinfo {year} {2018})}\BibitemShut {NoStop}%
\bibitem [{\citenamefont {You}\ \emph {et~al.}(2019)\citenamefont {You}, \citenamefont {Pearce}, \citenamefont {Sengupta},\ and\ \citenamefont {Giomi}}]{you2019}%
  \BibitemOpen
  \bibfield  {author} {\bibinfo {author} {\bibfnamefont {Z.}~\bibnamefont {You}}, \bibinfo {author} {\bibfnamefont {D.~J.}\ \bibnamefont {Pearce}}, \bibinfo {author} {\bibfnamefont {A.}~\bibnamefont {Sengupta}},\ and\ \bibinfo {author} {\bibfnamefont {L.}~\bibnamefont {Giomi}},\ }\href {https://doi.org/10.1103/PhysRevLett.123.178001} {\bibfield  {journal} {\bibinfo  {journal} {Phys. Rev. Lett.}\ }\textbf {\bibinfo {volume} {123}},\ \bibinfo {pages} {178001} (\bibinfo {year} {2019})}\BibitemShut {NoStop}%
\bibitem [{\citenamefont {Hosoda}\ \emph {et~al.}(2011)\citenamefont {Hosoda}, \citenamefont {Matsuura}, \citenamefont {Suzuki},\ and\ \citenamefont {Yomo}}]{hosoda2011}%
  \BibitemOpen
  \bibfield  {author} {\bibinfo {author} {\bibfnamefont {K.}~\bibnamefont {Hosoda}}, \bibinfo {author} {\bibfnamefont {T.}~\bibnamefont {Matsuura}}, \bibinfo {author} {\bibfnamefont {H.}~\bibnamefont {Suzuki}},\ and\ \bibinfo {author} {\bibfnamefont {T.}~\bibnamefont {Yomo}},\ }\href {https://doi.org/10.1103/PhysRevE.83.031118} {\bibfield  {journal} {\bibinfo  {journal} {Phys. Rev. E}\ }\textbf {\bibinfo {volume} {83}},\ \bibinfo {pages} {031118} (\bibinfo {year} {2011})}\BibitemShut {NoStop}%
\bibitem [{\citenamefont {Osella}\ \emph {et~al.}(2014)\citenamefont {Osella}, \citenamefont {Nugent},\ and\ \citenamefont {Cosentino~Lagomarsino}}]{osella2014concerted}%
  \BibitemOpen
  \bibfield  {author} {\bibinfo {author} {\bibfnamefont {M.}~\bibnamefont {Osella}}, \bibinfo {author} {\bibfnamefont {E.}~\bibnamefont {Nugent}},\ and\ \bibinfo {author} {\bibfnamefont {M.}~\bibnamefont {Cosentino~Lagomarsino}},\ }\href {https://doi.org/10.1073/pnas.1313715111} {\bibfield  {journal} {\bibinfo  {journal} {Proc. Natl. Acad. Sci. USA}\ }\textbf {\bibinfo {volume} {111}},\ \bibinfo {pages} {3431} (\bibinfo {year} {2014})}\BibitemShut {NoStop}%
\bibitem [{\citenamefont {Grilli}\ \emph {et~al.}(2017)\citenamefont {Grilli}, \citenamefont {Osella}, \citenamefont {Kennard},\ and\ \citenamefont {Lagomarsino}}]{grilli2017relevant}%
  \BibitemOpen
  \bibfield  {author} {\bibinfo {author} {\bibfnamefont {J.}~\bibnamefont {Grilli}}, \bibinfo {author} {\bibfnamefont {M.}~\bibnamefont {Osella}}, \bibinfo {author} {\bibfnamefont {A.~S.}\ \bibnamefont {Kennard}},\ and\ \bibinfo {author} {\bibfnamefont {M.~C.}\ \bibnamefont {Lagomarsino}},\ }\href {https://doi.org/10.1103/PhysRevE.95.032411} {\bibfield  {journal} {\bibinfo  {journal} {Phys. Rev. E}\ }\textbf {\bibinfo {volume} {95}},\ \bibinfo {pages} {032411} (\bibinfo {year} {2017})}\BibitemShut {NoStop}%
\bibitem [{\citenamefont {Nieto}\ \emph {et~al.}(2020)\citenamefont {Nieto}, \citenamefont {Arias-Castro}, \citenamefont {S{\'a}nchez}, \citenamefont {Vargas-Garc{\'\i}a},\ and\ \citenamefont {Pedraza}}]{nieto2020unification}%
  \BibitemOpen
  \bibfield  {author} {\bibinfo {author} {\bibfnamefont {C.}~\bibnamefont {Nieto}}, \bibinfo {author} {\bibfnamefont {J.}~\bibnamefont {Arias-Castro}}, \bibinfo {author} {\bibfnamefont {C.}~\bibnamefont {S{\'a}nchez}}, \bibinfo {author} {\bibfnamefont {C.}~\bibnamefont {Vargas-Garc{\'\i}a}},\ and\ \bibinfo {author} {\bibfnamefont {J.~M.}\ \bibnamefont {Pedraza}},\ }\href {https://doi.org/10.1103/PhysRevE.101.022401} {\bibfield  {journal} {\bibinfo  {journal} {Phys. Rev. E}\ }\textbf {\bibinfo {volume} {101}},\ \bibinfo {pages} {022401} (\bibinfo {year} {2020})}\BibitemShut {NoStop}%
\bibitem [{\citenamefont {Facchetti}\ \emph {et~al.}(2019)\citenamefont {Facchetti}, \citenamefont {Knapp}, \citenamefont {Chang},\ and\ \citenamefont {Howard}}]{facchetti2019reassessment}%
  \BibitemOpen
  \bibfield  {author} {\bibinfo {author} {\bibfnamefont {G.}~\bibnamefont {Facchetti}}, \bibinfo {author} {\bibfnamefont {B.}~\bibnamefont {Knapp}}, \bibinfo {author} {\bibfnamefont {F.}~\bibnamefont {Chang}},\ and\ \bibinfo {author} {\bibfnamefont {M.}~\bibnamefont {Howard}},\ }\href {https://doi.org/10.1016/j.bpj.2019.09.031} {\bibfield  {journal} {\bibinfo  {journal} {Biophys. J.}\ }\textbf {\bibinfo {volume} {117}},\ \bibinfo {pages} {1728} (\bibinfo {year} {2019})}\BibitemShut {NoStop}%
\bibitem [{\citenamefont {Le~Treut}\ \emph {et~al.}(2021)\citenamefont {Le~Treut}, \citenamefont {Si}, \citenamefont {Li},\ and\ \citenamefont {Jun}}]{leTreut2021quantitative}%
  \BibitemOpen
  \bibfield  {author} {\bibinfo {author} {\bibfnamefont {G.}~\bibnamefont {Le~Treut}}, \bibinfo {author} {\bibfnamefont {F.}~\bibnamefont {Si}}, \bibinfo {author} {\bibfnamefont {D.}~\bibnamefont {Li}},\ and\ \bibinfo {author} {\bibfnamefont {S.}~\bibnamefont {Jun}},\ }\href {https://doi.org/10.3389/fmicb.2021.721899} {\bibfield  {journal} {\bibinfo  {journal} {Front. Microbiol.}\ }\textbf {\bibinfo {volume} {12}},\ \bibinfo {pages} {721899} (\bibinfo {year} {2021})}\BibitemShut {NoStop}%
\bibitem [{\citenamefont {Luo}\ \emph {et~al.}(2023)\citenamefont {Luo}, \citenamefont {Bai},\ and\ \citenamefont {Fu}}]{luo2023stochastic}%
  \BibitemOpen
  \bibfield  {author} {\bibinfo {author} {\bibfnamefont {L.}~\bibnamefont {Luo}}, \bibinfo {author} {\bibfnamefont {Y.}~\bibnamefont {Bai}},\ and\ \bibinfo {author} {\bibfnamefont {X.}~\bibnamefont {Fu}},\ }\href {https://doi.org/10.1103/PhysRevResearch.5.013173} {\bibfield  {journal} {\bibinfo  {journal} {Phys. Rev. Res.}\ }\textbf {\bibinfo {volume} {5}},\ \bibinfo {pages} {013173} (\bibinfo {year} {2023})}\BibitemShut {NoStop}%
\bibitem [{\citenamefont {Genthon}\ and\ \citenamefont {Thomas}()}]{genthon2026cell}%
  \BibitemOpen
  \bibfield  {author} {\bibinfo {author} {\bibfnamefont {A.}~\bibnamefont {Genthon}}\ and\ \bibinfo {author} {\bibfnamefont {P.}~\bibnamefont {Thomas}},\ }\href@noop {} {\bibinfo {title} {Cell size control in bacteria is modulated through extrinsic noise, single-cell- and population-growth}},\ \Eprint {https://arxiv.org/abs/2601.05193} {arXiv:2601.05193 [q-bio.PE]} \BibitemShut {NoStop}%
\bibitem [{\citenamefont {Wallden}\ \emph {et~al.}(2016)\citenamefont {Wallden}, \citenamefont {Fange}, \citenamefont {Lundius}, \citenamefont {Baltekin},\ and\ \citenamefont {Elf}}]{wallden2016synchronization}%
  \BibitemOpen
  \bibfield  {author} {\bibinfo {author} {\bibfnamefont {M.}~\bibnamefont {Wallden}}, \bibinfo {author} {\bibfnamefont {D.}~\bibnamefont {Fange}}, \bibinfo {author} {\bibfnamefont {E.~G.}\ \bibnamefont {Lundius}}, \bibinfo {author} {\bibfnamefont {{\"O}.}~\bibnamefont {Baltekin}},\ and\ \bibinfo {author} {\bibfnamefont {J.}~\bibnamefont {Elf}},\ }\href {https://doi.org/10.1016/j.cell.2016.06.052} {\bibfield  {journal} {\bibinfo  {journal} {Cell}\ }\textbf {\bibinfo {volume} {166}},\ \bibinfo {pages} {729} (\bibinfo {year} {2016})}\BibitemShut {NoStop}%
\bibitem [{\citenamefont {Lee}\ \emph {et~al.}(2019)\citenamefont {Lee}, \citenamefont {Wu},\ and\ \citenamefont {Errington}}]{lee2019microfluidic}%
  \BibitemOpen
  \bibfield  {author} {\bibinfo {author} {\bibfnamefont {S.}~\bibnamefont {Lee}}, \bibinfo {author} {\bibfnamefont {L.~J.}\ \bibnamefont {Wu}},\ and\ \bibinfo {author} {\bibfnamefont {J.}~\bibnamefont {Errington}},\ }\href {https://doi.org/10.1002/mbo3.876} {\bibfield  {journal} {\bibinfo  {journal} {MicrobiologyOpen}\ }\textbf {\bibinfo {volume} {8}},\ \bibinfo {pages} {e876} (\bibinfo {year} {2019})}\BibitemShut {NoStop}%
\bibitem [{\citenamefont {Nieto}\ \emph {et~al.}(2024)\citenamefont {Nieto}, \citenamefont {Vargas-Garc{\'\i}a}, \citenamefont {Pedraza},\ and\ \citenamefont {Singh}}]{nieto2024mechanisms}%
  \BibitemOpen
  \bibfield  {author} {\bibinfo {author} {\bibfnamefont {C.}~\bibnamefont {Nieto}}, \bibinfo {author} {\bibfnamefont {C.~A.}\ \bibnamefont {Vargas-Garc{\'\i}a}}, \bibinfo {author} {\bibfnamefont {J.~M.}\ \bibnamefont {Pedraza}},\ and\ \bibinfo {author} {\bibfnamefont {A.}~\bibnamefont {Singh}},\ }\href {https://doi.org/10.1038/s41540-024-00383-z} {\bibfield  {journal} {\bibinfo  {journal} {npj Syst. Biol. Appl.}\ }\textbf {\bibinfo {volume} {10}},\ \bibinfo {pages} {61} (\bibinfo {year} {2024})}\BibitemShut {NoStop}%
\bibitem [{\citenamefont {Panlilio}\ \emph {et~al.}(2021)\citenamefont {Panlilio}, \citenamefont {Grilli}, \citenamefont {Tallarico}, \citenamefont {Iuliani}, \citenamefont {Sclavi}, \citenamefont {Cicuta},\ and\ \citenamefont {Cosentino~Lagomarsino}}]{panlilio2021threshold}%
  \BibitemOpen
  \bibfield  {author} {\bibinfo {author} {\bibfnamefont {M.}~\bibnamefont {Panlilio}}, \bibinfo {author} {\bibfnamefont {J.}~\bibnamefont {Grilli}}, \bibinfo {author} {\bibfnamefont {G.}~\bibnamefont {Tallarico}}, \bibinfo {author} {\bibfnamefont {I.}~\bibnamefont {Iuliani}}, \bibinfo {author} {\bibfnamefont {B.}~\bibnamefont {Sclavi}}, \bibinfo {author} {\bibfnamefont {P.}~\bibnamefont {Cicuta}},\ and\ \bibinfo {author} {\bibfnamefont {M.}~\bibnamefont {Cosentino~Lagomarsino}},\ }\href {https://doi.org/10.1073/pnas.2016391118} {\bibfield  {journal} {\bibinfo  {journal} {Proc. Natl. Acad. Sci. USA}\ }\textbf {\bibinfo {volume} {118}},\ \bibinfo {pages} {e2016391118} (\bibinfo {year} {2021})}\BibitemShut {NoStop}%
\bibitem [{\citenamefont {Shi}\ \emph {et~al.}(2021)\citenamefont {Shi}, \citenamefont {Hu}, \citenamefont {Odermatt}, \citenamefont {Gonzalez}, \citenamefont {Zhang}, \citenamefont {Elias}, \citenamefont {Chang},\ and\ \citenamefont {Huang}}]{shi2021precise}%
  \BibitemOpen
  \bibfield  {author} {\bibinfo {author} {\bibfnamefont {H.}~\bibnamefont {Shi}}, \bibinfo {author} {\bibfnamefont {Y.}~\bibnamefont {Hu}}, \bibinfo {author} {\bibfnamefont {P.~D.}\ \bibnamefont {Odermatt}}, \bibinfo {author} {\bibfnamefont {C.~G.}\ \bibnamefont {Gonzalez}}, \bibinfo {author} {\bibfnamefont {L.}~\bibnamefont {Zhang}}, \bibinfo {author} {\bibfnamefont {J.~E.}\ \bibnamefont {Elias}}, \bibinfo {author} {\bibfnamefont {F.}~\bibnamefont {Chang}},\ and\ \bibinfo {author} {\bibfnamefont {K.~C.}\ \bibnamefont {Huang}},\ }\href {https://doi.org/10.1038/s41467-021-22092-5} {\bibfield  {journal} {\bibinfo  {journal} {Nat. Commun.}\ }\textbf {\bibinfo {volume} {12}},\ \bibinfo {pages} {1975} (\bibinfo {year} {2021})}\BibitemShut {NoStop}%
\bibitem [{\citenamefont {B{\"u}ke}\ \emph {et~al.}(2022)\citenamefont {B{\"u}ke}, \citenamefont {Grilli}, \citenamefont {Lagomarsino}, \citenamefont {Bokinsky},\ and\ \citenamefont {Tans}}]{buke2022ppgpp}%
  \BibitemOpen
  \bibfield  {author} {\bibinfo {author} {\bibfnamefont {F.}~\bibnamefont {B{\"u}ke}}, \bibinfo {author} {\bibfnamefont {J.}~\bibnamefont {Grilli}}, \bibinfo {author} {\bibfnamefont {M.~C.}\ \bibnamefont {Lagomarsino}}, \bibinfo {author} {\bibfnamefont {G.}~\bibnamefont {Bokinsky}},\ and\ \bibinfo {author} {\bibfnamefont {S.~J.}\ \bibnamefont {Tans}},\ }\href {https://doi.org/10.1016/j.cub.2021.12.033} {\bibfield  {journal} {\bibinfo  {journal} {Curr. Biol.}\ }\textbf {\bibinfo {volume} {32}},\ \bibinfo {pages} {870} (\bibinfo {year} {2022})}\BibitemShut {NoStop}%
\bibitem [{\citenamefont {Pahujani}\ \emph {et~al.}(2026)\citenamefont {Pahujani}, \citenamefont {Zhang}, \citenamefont {Stetter},\ and\ \citenamefont {Krug}}]{pahujani2026adaptive}%
  \BibitemOpen
  \bibfield  {author} {\bibinfo {author} {\bibfnamefont {S.}~\bibnamefont {Pahujani}}, \bibinfo {author} {\bibfnamefont {Y.}~\bibnamefont {Zhang}}, \bibinfo {author} {\bibfnamefont {M.~G.}\ \bibnamefont {Stetter}},\ and\ \bibinfo {author} {\bibfnamefont {J.}~\bibnamefont {Krug}},\ }\href {https://doi.org/10.1093/genetics/iyag168} {\bibfield  {journal} {\bibinfo  {journal} {Genetics}\ }\textbf {\bibinfo {volume} {234}},\ \bibinfo {pages} {iyag168} (\bibinfo {year} {2026})}\BibitemShut {NoStop}%
\bibitem [{\citenamefont {Jin}\ and\ \citenamefont {Sengupta}(2025)}]{dataset}%
  \BibitemOpen
  \bibfield  {author} {\bibinfo {author} {\bibfnamefont {C.}~\bibnamefont {Jin}}\ and\ \bibinfo {author} {\bibfnamefont {A.}~\bibnamefont {Sengupta}},\ }\href {https://doi.org/10.5281/zenodo.17768423} {\bibinfo {title} {Mechanical confinement induces a transient timer-like shift in bacterial size control, version 2}},\ \bibinfo {howpublished} {Zenodo} (\bibinfo {year} {2025})\BibitemShut {NoStop}%
\end{thebibliography}%

\end{document}